\documentclass{article}
\usepackage{graphicx}
\usepackage[margin=1.5in]{geometry}
\usepackage{enumerate}
\usepackage{color}
\usepackage{verbatim}
\usepackage{imakeidx}
 
\makeindex

\title{Bio-inspired underwater propulsors}
\author{Tyler Van Buren, Daniel Floryan, and Alexander J. Smits}
\date{\today}

\begin{document}
 
\maketitle

\noindent In recent years, there has been considerable interest in developing novel underwater vehicles that use propulsion systems inspired by biology \cite{colgate2004mechanics,bandyopadhyay2005trends}. Such vehicles have the potential to uncover new mission capabilities and improve maneuverability\index{maneuverability}, efficiency\index{swimming efficiency}, and speed \cite{fish2003conceptual,fishbiomimetic}. Here we will explore the physical mechanisms that govern the performance\textemdash especially swimming speed\index{swimming speed} and efficiency\textemdash of propulsive techniques inspired by biology.  We will also show we can translate the understanding we have gained from biology to the design of a new generation of underwater vehicles.

Many aquatic animals are capable of great speed, high efficiency, and rapid maneuvering.  Engineers have been able to mimic their behavior by constructing robotic imitations, with some considerable success.  One of the best-known examples is Robotuna, an eight link tendon- and pulley-driven, whose external shape has the form of a bluefin tuna, capable of emulating the swimming motion of a live tuna\index{swimming animals!tuna} \cite{triantafyllou1995efficient}.  This project evolved into the Ghostswimmer, a prototype Navy vehicle that swims by manipulating its dorsal (back), pectoral (chest), and caudal (tail) fins \cite{rufo2011ghostswimmer}.  However, many features of biology do not exist for the purpose of swimming alone and could exist for survivability or reproductive purposes. As a design paradigm, therefore, it may be better to abandon biomimetic designs for ones that are inspired by biology but not constrained by it.  

\begin{figure}
\centering
\includegraphics[width=1\textwidth]{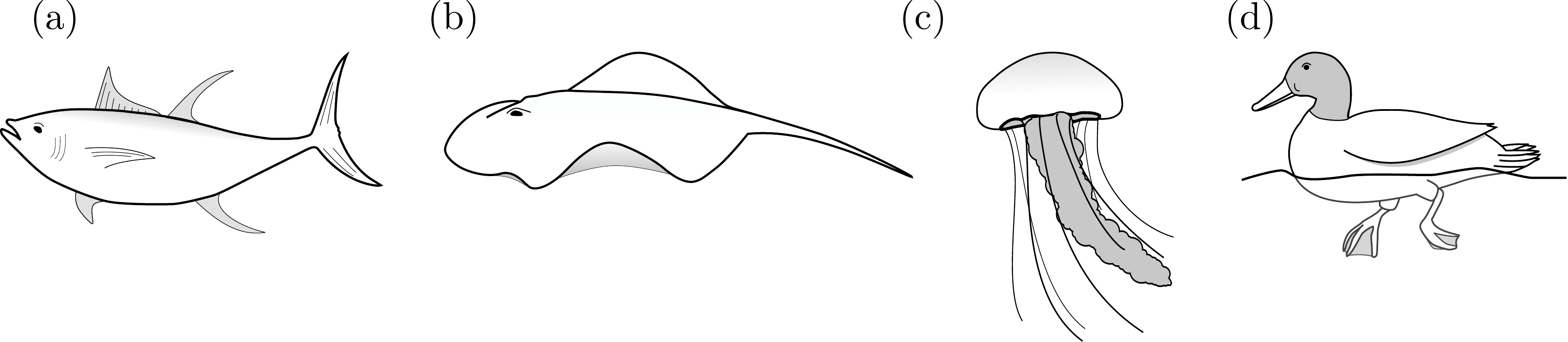}
\caption{Examples of four swimming types: (a) oscillatory - tuna\index{swimming animals!tuna}, (b) undulatory - ray\index{swimming animals!ray}, (c) pulsatile jet\index{jet} - jellyfish\index{swimming animals!jellyfish}, and (d) drag-based\index{thrust!drag-based} - duck\index{swimming animals!duck}\index{swimming type}.}
\label{fig:swimmerType}
\end{figure}

In general, we can identify  four major types of swimmers, with examples illustrated in figure \ref{fig:swimmerType}:

\begin{itemize}
\item[] {\bf Oscillatory}: these animals propel themselves primarily using a semi-rigid caudal fin\index{caudal fin} or fluke\index{fluke} that is oscillated periodically. Examples include salmon\index{swimming animals!salmon}, tuna\index{swimming animals!tuna}, and dolphin\index{swimming animals!dolphin}.\index{swimming type!oscillatory}
\item[] {\bf Undulatory}: these animals utilize a traveling wave along their body or propulsive fins to push fluid backward. Examples include eels\index{swimming animals!eel}, lampreys\index{swimming animals!lamprey},  and rays\index{swimming animals!ray}.  \index{swimming type!undulatory}
\item[] {\bf Pulsatile}: these animals periodically ``inhale'' a volume of water and then discharge it impulsively as a jet\index{jet}, producing thrust\index{thrust} in the direction opposite the jet. Examples include jellyfish\index{swimming animals!jellyfish}, squid\index{swimming animals!squid}, and some mollusks\index{swimming animals!mollusk}.\index{swimming type!pulsatile}
\item[] {\bf Drag-based}:\index{thrust!drag-based} these animals force a bluff body\index{bluff body} such as a rigid flipper through the water to generate thrust by reaction. Examples include humans\index{swimming animals!human}, turtles\index{swimming animals!turtle}, and ducks\index{swimming animals!duck}.\index{swimming type!drag-based}
\end{itemize}\index{swimming type}

\noindent Some swimmers use more than one swimming type. For example, the sea turtle\index{swimming animals!turtle} has been reported to use the \emph{power stroke}\index{power stroke} \cite{lutz2002}, where the first half of the stroke is drag-based\index{thrust!drag-based} and the second half is oscillatory. In this chapter, we will treat the different swimming types in isolation, though we encourage readers to think about possible designs that could incorporate positive aspects of each.

\begin{figure}
\centering
\includegraphics[width=1\textwidth]{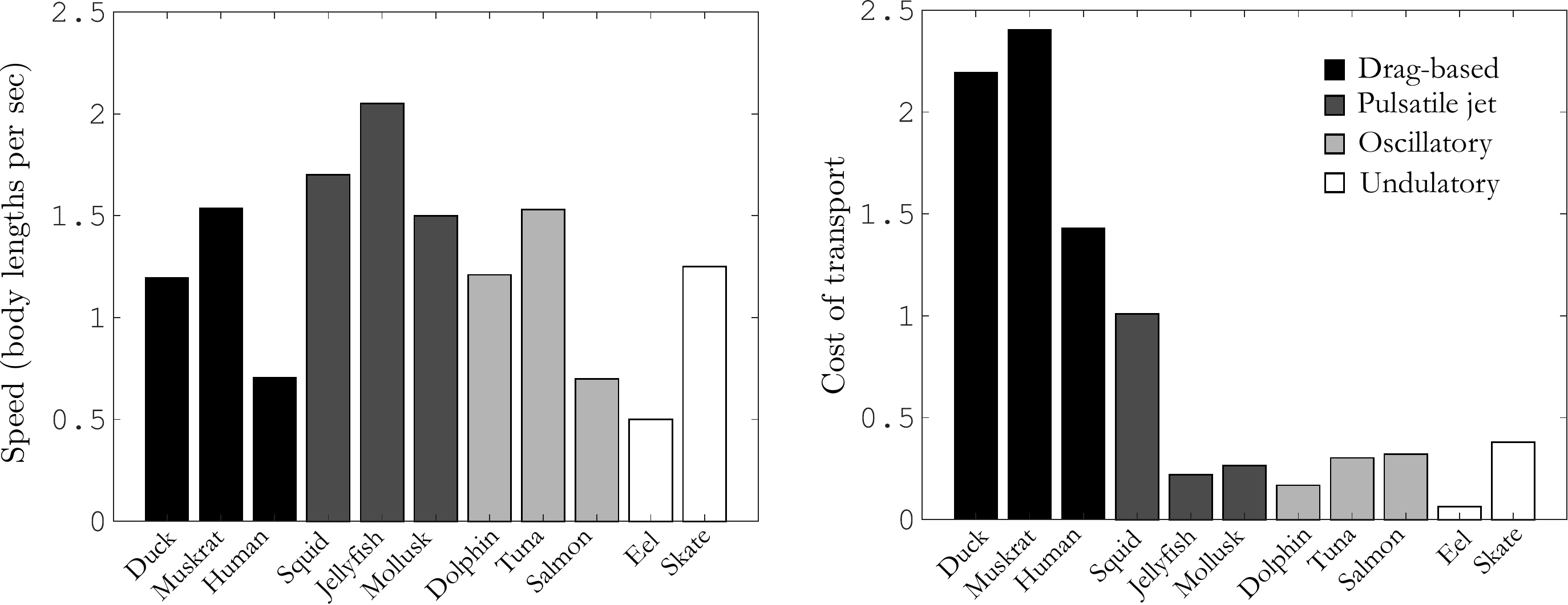}
\caption{Relative swimming speed and cost of transport\index{cost of transport} for oscillatory, undulatory, pulsatile jet\index{jet}, and drag-based swimmers. Data summarized from \cite{fish2000}.}
\label{fig:swimmerAbility}
\end{figure}

It is not surprising that different swimmers show differences in their performance characteristics. Typical swimming speed\index{swimming speed} (normalized by body length) and cost of transport\index{cost of transport} of the four types of swimmers are shown in figure \ref{fig:swimmerAbility}.  The cost of transport\index{cost of transport} quantifies the energy efficiency\index{swimming efficiency} of transporting an animal or vehicle from one place to another.  In biology, it is often expressed as the distance traveled per unit energy cost (similar to miles per gallon).  Despite large differences in size and swimming mechanisms, we see that most organisms swim between 0.5 and 1.5 body lengths per second.  This is a typical cruising speed, and the maximum swimming speed\index{swimming speed} can be very different among different swimmer types. However, drag-based swimmers\index{thrust!drag-based} have a notably higher cost of transport\index{cost of transport} than the others. This makes sense, as most drag-based swimmers do not necessarily solely live in water, and many have evolved to also walk or fly. 

We now examine the basic mechanisms that these different types of swimmers employ, as they might be implemented on a vehicle.

\subsection{Mechanics of underwater propulsion}\index{underwater propulsion}

In steady swimming, where there is no acceleration or deceleration, the thrust\index{thrust} produced by the propulsive system is balanced exactly by the drag\index{drag} on the vehicle, in the time-average.  For underwater vehicles\index{underwater vehicle}, the drag force has two major components: the friction drag due to the viscous shear stresses acting on the surface of the vehicle, and the pressure, or form, drag\index{drag!pressure/form drag} due to the pressure losses in the wake\index{wake}. For streamlined vehicles, such as those shaped like fish, the viscous drag\index{drag!viscous drag} component tends to dominate, whereas for bluff bodies, exemplified by more boxy shapes, the form drag dominates. An important parameter is the Reynolds number\index{Reynolds number} $Re$, which is a measure of the importance of inertial forces to viscous forces, and is defined by $Re=\rho U_\infty L/\mu$, where $U_\infty$ is the speed, $L$ is a characteristic dimension of the vehicle such as its length, and $\rho$ and $\mu$ are the density and viscosity of the fluid, in our case fresh or salt water.  At large Reynolds numbers\index{Reynolds number}, typical of most fish and all underwater vehicles, the form drag is almost independent of Reynolds number, but the viscous drag always remains a function of Reynolds number.

\begin{figure}
\centering
\includegraphics[width=0.75\textwidth]{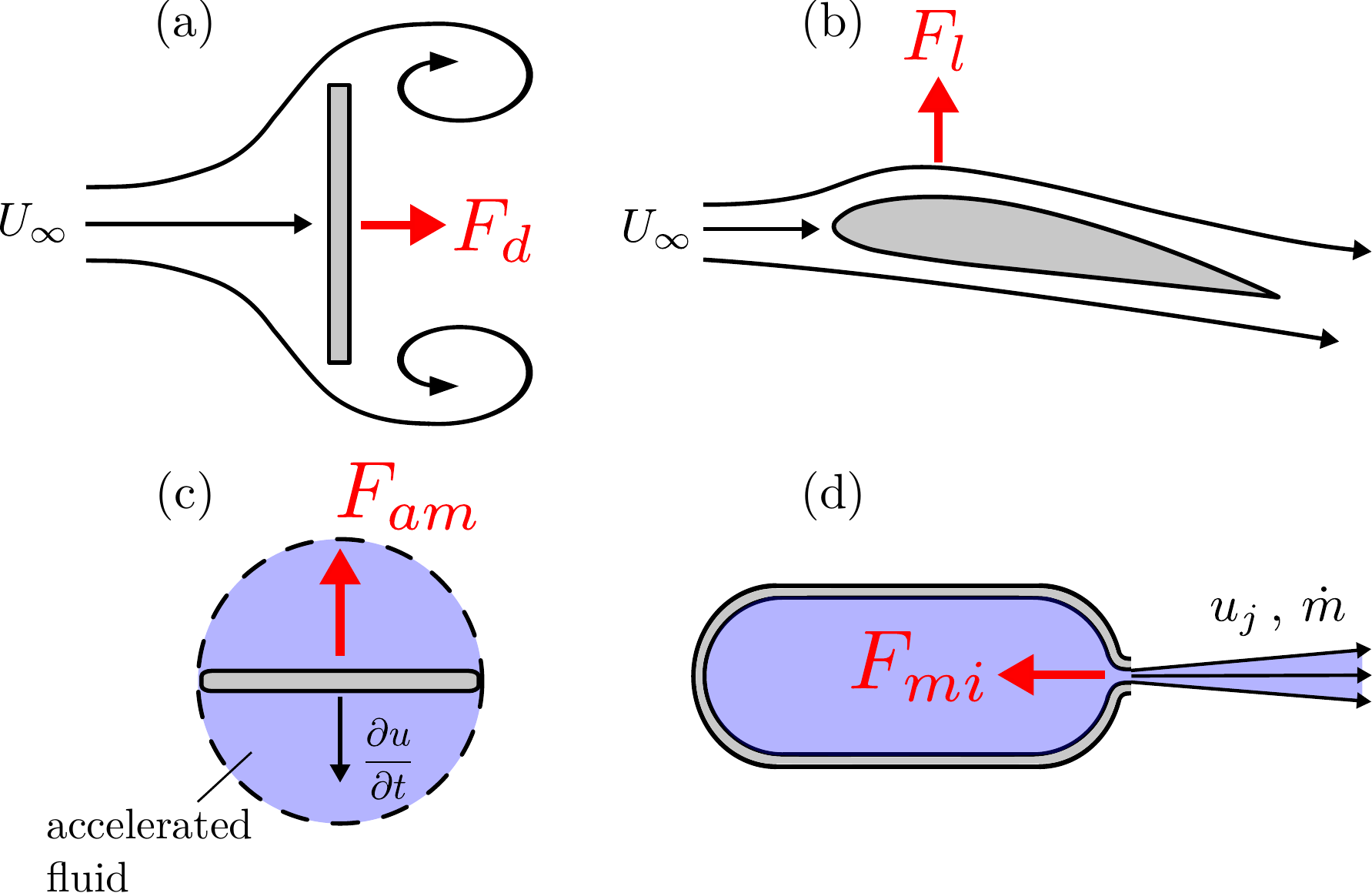}
\caption{Typical thrust generation mechanisms: (a) drag-based\index{thrust!drag-based}, (b) lift-based\index{thrust!lift-based}, (c) added mass\index{added mass}, and (d) momentum injection\index{momentum injection}.}
\label{fig:thrustType} 
\end{figure}

For thrust, most modern human-designed propulsors utilize some sort of continuous rotation (think propellers), which is not a motion natural to biology.  Fish and mammals such as dolphins\index{swimming animals!dolphin} and whales use fins and flukes\index{fluke} to propel themselves in combined pitching\index{pitch} and heaving\index{heave} motions, turtles\index{swimming animals!turtle} use a paddling motion, while squid\index{swimming animals!squid} eject jets\index{jet} of fluid.  We find that there are four major sources of thrust\index{thrust}: (1) drag-based thrust\index{thrust!drag-based}, (2) lift-based thrust\index{thrust!lift-based}, (3) added mass\index{added mass} forces, and (4) momentum injection\index{momentum injection}, as illustrated in figure \ref{fig:thrustType}.  Drag is the force acting opposite to the direction of motion of the body, lift\index{lift} is the force produced normal to the direction of motion, added mass forces are due to the inertia of the water that is put in motion by the body, and momentum injection\index{momentum injection} is the force produced by jetting\index{jet} fluid from the body, as used by squid\index{swimming animals!squid} and jellyfish\index{swimming animals!jellyfish}.  These propulsion types can be, and often are, combined in practical systems, but first we consider them separately. 

Drag-based thrust\index{thrust!drag-based} actually uses the form drag\index{drag!pressure/form drag} experienced by a bluff body\index{bluff body} to generate thrust.  Human\index{swimming animals!human} swimmers make extensive use of drag-based thrust\index{thrust!drag-based}.  For example, a swimmer doing the breaststroke will spread her hands, push water rearwards, and so by action-reaction propels herself forward.  The (steady) drag-based thrust $F_d$ is given by
\begin{equation}
F_{d}=\frac{1}{2}\rho U_\infty^2\,A\,C_D,
\label{eqn:drag}
\end{equation}
where $A$ is the frontal area of the body, $U_\infty$ is the speed, and the drag coefficient\index{coefficient!drag} $C_D$ is defined by the body shape (see figure~\ref{fig:thrustType}a).  The drag coefficient\index{coefficient!drag} of a hand is about 1, so it takes about 10 N of force to move a hand at 1 m/s through water.

As to lift-based thrust\index{thrust!lift-based}, we noted that lift\index{lift} was defined as the force acting normal to the direction of motion.  For a simple airfoil\index{foil!airfoil} or hydrofoil\index{foil!hydrofoil} in steady motion,  the lift\index{lift} force is given by
\begin{equation}
F_{l}=\frac{1}{2}\rho U_\infty^2\, A\,C_L,
\end{equation}
where $C_L$ is the lift\index{lift} coefficient\index{coefficient!lift}, which depends on the shape of the object and the angle of attack\index{angle of attack} $\alpha$ (see figure~\ref{fig:thrustType}b).  For a thin hydrofoil\index{foil!hydrofoil}, $C_L$ depends only on the angle of attack\index{angle of attack} ($=2 \pi \alpha$).  Because the lift\index{lift} is always perpendicular to the body motion, getting thrust from lift\index{lift} requires the propulsor to move laterally as well as translationally, so that the effective (local) velocity seen by the propulsor is such that a portion of the lift acts in a direction that produces thrust (that is, a forward velocity). This is sometimes referred to as the ``Knoller-Betz effect''.

Thrust due to added mass\index{added mass} may be illustrated by considering the differences between suddenly moving your hand in air, as compared to doing it in water.  In both cases, there would be a drag\index{drag} force, as described above, due to form drag\index{drag!pressure/form drag}, but there is an additional force required to move the surrounding fluid.  This ``added mass''\index{added mass} force $F_{am}$ depends on the mass of fluid put into motion, as well as the acceleration required (Newton's Second Law), as in
\begin{equation}
F_{am}=m_f\frac{\partial u}{\partial t},
\end{equation}
where $m_f$ is the fluid mass and $\partial u / \partial t$ is the fluid acceleration. It is relatively small in air but much larger in water, because the density of water is about 800 times that of air.  The reaction force generated by this impulsive motion in water can generate considerable thrust.  The actual force can be difficult to assess accurately because it is difficult to estimate precisely the volume of fluid that is accelerated by a particular motion, but for simple motions and shapes, some reasonable estimates can be given. There is often a misconception that added mass is more important in water than in air because water is heavier, but all of these forces scale with the fluid density, thus the weight of the fluid does not dictate the relative importance of thrust generation mechanisms.

Lastly, we consider the direct injection of momentum into the surrounding fluid, most commonly via a jet\index{jet}.  If you were to inflate a balloon, then release it before tying the bottom opening, it would erratically dart around the room while deflating. This is because the pressure inside the balloon is released through the small opening, creating a  jet\index{jet} of air.  This released air has momentum, and so it creates a reaction force that causes the balloon to move.  Similarly, if you were to let go of a running hose it would whip around dramatically.  In the steady case, the momentum of the jet\index{jet} governs the force generated according to
\begin{equation}
F_{mi}=\dot{m}v_j,
\end{equation}
where $\dot{m}$ is the mass flow rate through the jet\index{jet} opening and $v_j$ is the jet\index{jet} velocity. When properly harnessed, this force can be utilized as a controlled thrust generation mechanism. 

We have thus far identified four possible sources of thrust that are important for underwater propulsion\index{underwater propulsion}.  In addition, we need to consider the efficiency, which is typically defined by 
\begin{equation}
\eta=\frac{F_x\,U_\infty}{\mathcal{P}},
\end{equation}
where $\mathcal{P}$ is the power input into the propulsor, $F_x$ is the thrust generated by the propulsor, and $U_\infty$ is the velocity of the vehicle. This definition is termed the `Froude efficiency'\index{efficiency!Froude} and it is the fraction of power input into the propulsor that is used to propel the vehicle forward, ignoring the question of how efficiently the input power was generated to begin with. We emphasize that the numerator is the work that the system needs to do against the drag\index{drag} experienced by the vehicle, that is, for steady swimming $F_x$ balances the vehicle drag such that the vehicle moves at a constant speed. The denominator is the power input into the propulsor, or the power that the propulsor has at its disposal in order to do work against the drag experienced by the vehicle. The form of the numerator is as written above for all steadily swimming vehicles, whereas the form of the input power in the denominator differs for different modes of propulsion. 

With these basic concepts in place, we can now examine the traits of different swimmers (oscillatory, undulatory, pulsatile, and drag-based\index{thrust!drag-based}) to identify elements that could inspire propulsor design, with a special focus on thrust and efficiency. Due to the similarities in thrust production mechanisms and wake\index{wake} characteristics, we will combine oscillatory and undulatory swimmers into a single section. For each swimming style, we will present our current understanding of the fundamental mechanisms that govern propulsive performance. Then, we will propose a simplified robotic swimmer concept for each swimming type where we use aquatic swimmers as inspiration, without constraining ourselves to mimic biology.

\subsection{Oscillatory and undulatory}\index{swimming type!oscillatory}\index{swimming type!undulatory}

\begin{figure}
\centering
\includegraphics[width=0.75\textwidth]{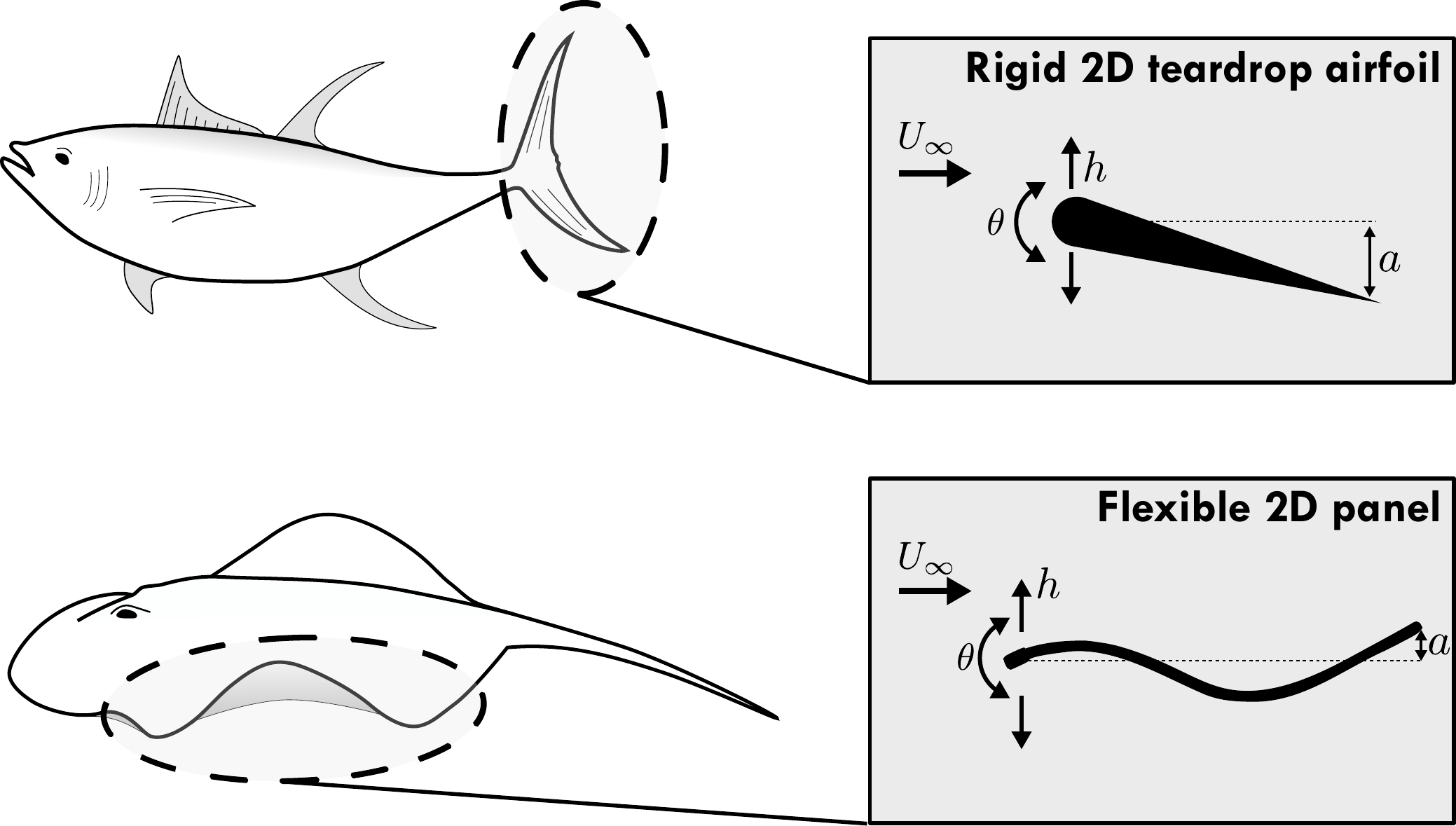}
\caption{Example of how one could simplify an oscillatory body/caudal fin\index{caudal fin} (top) and undulatory median/paired fins (bottom) swimmers. Motion inputs: heave\index{heave} $h(t)$ and pitch\index{pitch} $\theta(t)$, and output: trailing edge amplitude $a(t)$.}
\label{fig:propulsorSimple}
\end{figure}

Oscillatory and undulatory swimmers make up a large fraction of the aquatic life synonymous with high swimming speed\index{swimming speed} and efficiency\index{efficiency!swimming}, and so they have become the focus of propulsion inspiration. These swimmers use lateral motion of their propulsive surfaces to generate thrust. \emph{Thunniform}\index{thunniform} swimmers, such as tuna\index{swimming animals!tuna}, tend to be more oscillatory swimmers--- the primary propulsion is achieved using a flapping motion of their caudal fin\index{caudal fin} or fluke\index{fluke} and the body is fairly rigid. \emph{Anguilliform}\index{anguilliform} swimmers such as eels\index{swimming animals!eel} and snakes are more undulatory--- the body itself is the primary propulsor.  Rays\index{swimming animals!ray} also use undulatory motion, but they use their elongated pectoral fins\index{pectoral fin} instead of a caudal fin.

Figure \ref{fig:propulsorSimple} shows examples of how to model an oscillatory or undulatory swimmer. For example, an oscillatory swimmer's propulsor can be modeled using a pitching\index{pitch} and heaving\index{heave} foil\index{foil!hydrofoil}, or an undulatory swimmer's propulsor could be modeled as a flexible panel.  Although these simple models may not look much like fish, they can reveal much of the underlying physics and provide simple designs for possible application to underwater vehicles\index{underwater vehicle}.

\subsubsection{\emph{Wake characteristics}}\label{S1}\index{wake}

\begin{figure}
\centering
\includegraphics[width=1\textwidth]{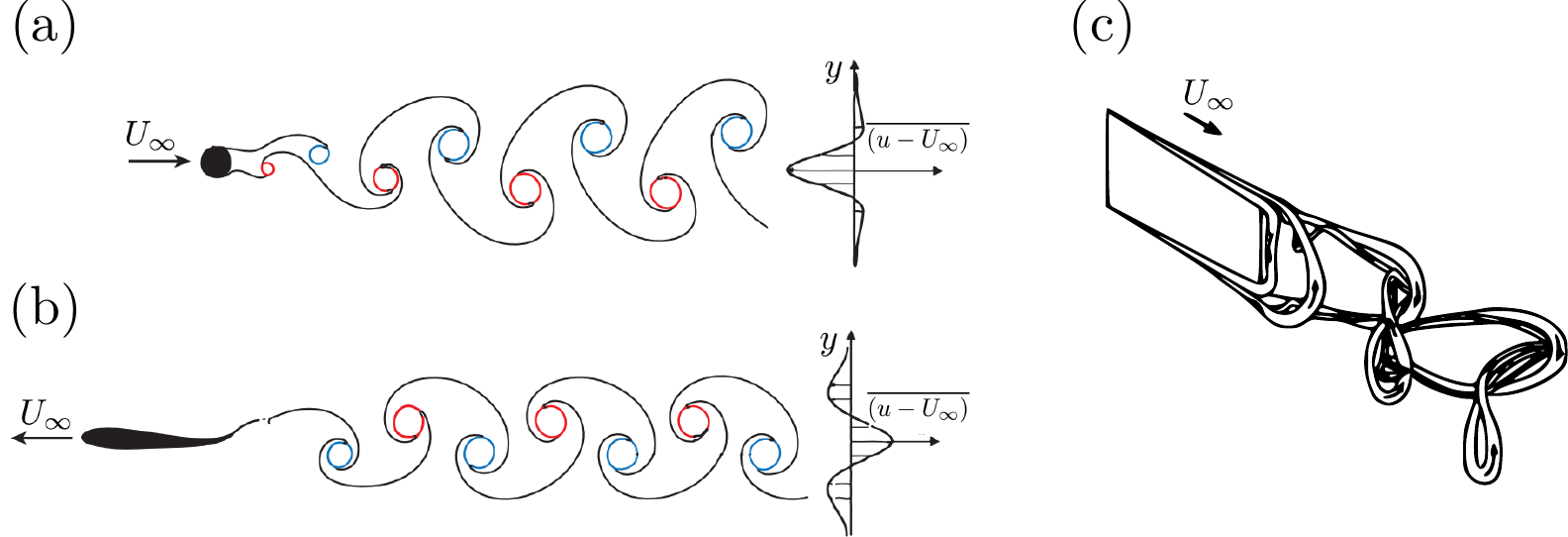}
\caption{Vortex\index{vortex} wake and time-averaged velocity profiles of a two-dimensional (a) drag-producing cylinder and (b) thrust-producing foil; (c) three-dimensional vortex skeleton of a pitching\index{pitch} panel. Taken from \cite{eloy2012, buchholz2006}.\index{flow visualization}}
\label{fig:wakes}
\end{figure}

The wake of a propulsor is like its footprint, in that it can reveal the physical mechanisms by which thrust is produced. For example, the flow over a stationary cylinder periodically separates, resulting in an oscillating drag force.  In the wake, this unsteadiness is evinced by a train of alternating sign vortices, known as a \emph{von K\'{a}rm\'{a}n} vortex\index{vortex!von K\'{a}rm\'{a}n} street, shown in figure \ref{fig:wakes}a.  If you were to measure the time-averaged velocity field downstream of the cylinder, you would see a mean momentum deficit in the wake, indicating that a net drag is acting on the body.  A similar wake is exhibited by a pitching\index{pitch} or heaving\index{heave} foil, but here the vortices are of the opposite sign, forming a \emph{reverse von K\'{a}rm\'{a}n} vortex\index{vortex!von K\'{a}rm\'{a}n (reverse)} street shown in figure \ref{fig:wakes}b.  This orientation results in a time-average velocity field that has a mean momentum excess in the wake, indicating thrust production. 

When the propulsor has a finite width, the wake becomes highly three-dimensional, as shown in figure \ref{fig:wakes}c.  The spanwise vortices seen in a two-dimensional propulsor become interconnected loops, and the interaction among these vortex\index{vortex} loops compresses the wake in the spanwise direction while spreading it in the panel-normal direction. Van Buren et al. \cite{VanBuren2017_1} altered the shape of the trailing edge to manipulate the vortex\index{vortex} dynamics of the wake. They found that they delayed or enhanced the wake compression and breakdown, impacting the thrust and efficiency of the propulsor.  Hence, it may be possible to control the performance of a propulsor by altering the vortex\index{vortex} structure in the wake. 

\subsubsection{\emph{Motion type}}

For our simplest model of a propulsor, we will neglect the influence of the main body of the vehicle, and consider a sinusoidally\index{motion!sinusoidal} oscillating rigid foil in isolation.  Its motion can be deconstructed into a time-varying pitching\index{pitch} component $\theta(t)$ (twisting about the leading edge), and heaving\index{heave} component $h(t)$ (pure lateral translation, or plunging), that together produce a trailing edge motion $a_t(t)$ (see figure \ref{fig:propulsorSimple}). The foil is assumed to be rectangular, with a chord length $c$ and a span $s$.

The thrust\index{thrust} generated by pitching\index{pitch} motion is due purely to added mass\index{added mass} forces, since the time-average lift\index{lift} force is zero \cite{Floryan2016}. We therefore expect the thrust to scale as the product of the streamwise component of the added mass\index{added mass} ($\sim \rho c^2 s$) and the acceleration ($\sim c \ddot \theta$), where the symbol $\sim$ denotes ``it varies as''. That is, 
$$F_x \sim \rho s c^3 \ddot \theta \theta, $$ 
and so the time-averaged thrust scales according to 
\begin{equation}
 \overline{F}_x \sim \rho s c^3 f^2 \theta_0^2 \ \approx \rho s c f^2 a^2,
 \label{pitch_scale}
\end{equation}
where $f$ is the frequency of oscillation and $a \approx c\theta_0$ is the trailing edge amplitude for small pitching\index{pitch} motions.

The thrust produced by heaving\index{heave} motions is primarily due to lift-based forces\index{thrust!lift-based}, and added mass\index{added mass} forces in the thrust direction are typically small \cite{Floryan2016}. Thus, we expect the thrust to scale as the streamwise component of the instantaneous lift\index{lift} force. That is, 
$$F_x \sim L ( {\dot h}/{U^*} ) $$,
where $L$ is the lift force, $\dot h$ is the heave\index{heave} velocity, and $U^*$ is the effective velocity seen by the foil. 
If we assume that the contribution to the lift is quasi-steady, and that the angle of attack\index{angle of attack} is small, so that $\alpha \approx {\dot h}/{U^*}$, then
$$F_x \sim {\textstyle \frac{1}{2}} \rho U^{*2} s c \, (2 \pi \alpha)  ( {\dot h}/{U^*} ) \sim  \pi \rho s c \, \dot h^2,$$
so that the mean thrust scales as 
\begin{equation}
\overline{F}_x \sim \rho s c f^2h_0^2 \ = \rho s c f^2a^2,
\label{heave_scale}
\end{equation}
where $a=h_0$ for heaving\index{heave} motions. Interestingly, at this level of approximation, the thrust generated through pitch\index{pitch} and heave\index{heave} scale similarly, and have no dependence on velocity (which has been confirmed experimentally by \cite{VanBuren2018}). A more detailed analysis is offered in Floryan et al. \cite{Floryan2016}, which includes unsteady and nonlinear effects.

Typically, the performance and motions are presented non-dimensionally. The thrust and power are then normalized using aerodynamic convention, and so we obtain the thrust and power coefficients\index{coefficient!thrust}\index{coefficient!power} as well as the resulting efficiency, defined by
\begin{equation}
C_T = \frac{F_x}{\frac{1}{2}\rho U_\infty^2sc},\qquad C_P=\frac{F_y\dot{h}+M_z\dot{\theta}}{\frac{1}{2}\rho U_\infty^3sc},\qquad
\eta = C_T/C_P,
\label{perfEQ}
\end{equation}
where $F_y$ is the side force and $M_z$ is the spanwise moment.  We also introduce the Strouhal number\index{Strouhal number},
\begin{equation}
St=\frac{2\,fa}{U_\infty},
\end{equation}
which can be thought of as a ratio of the wake width to the spacing between shed vortices, or as the ratio of the lateral velocity of the trailing edge to the freestream velocity.  Another parameter is the reduced frequency\index{reduced frequency}, defined by
\begin{equation}
f^*=\frac{fc}{U_\infty},
\end{equation}
which is the time it takes a particle to pass from the leading to the trailing edge of the propulsor compared to the period of its motion. Note that $f^*$ does not depend on the motion amplitude. 

\begin{figure}
\centering
\includegraphics[width=1\textwidth]{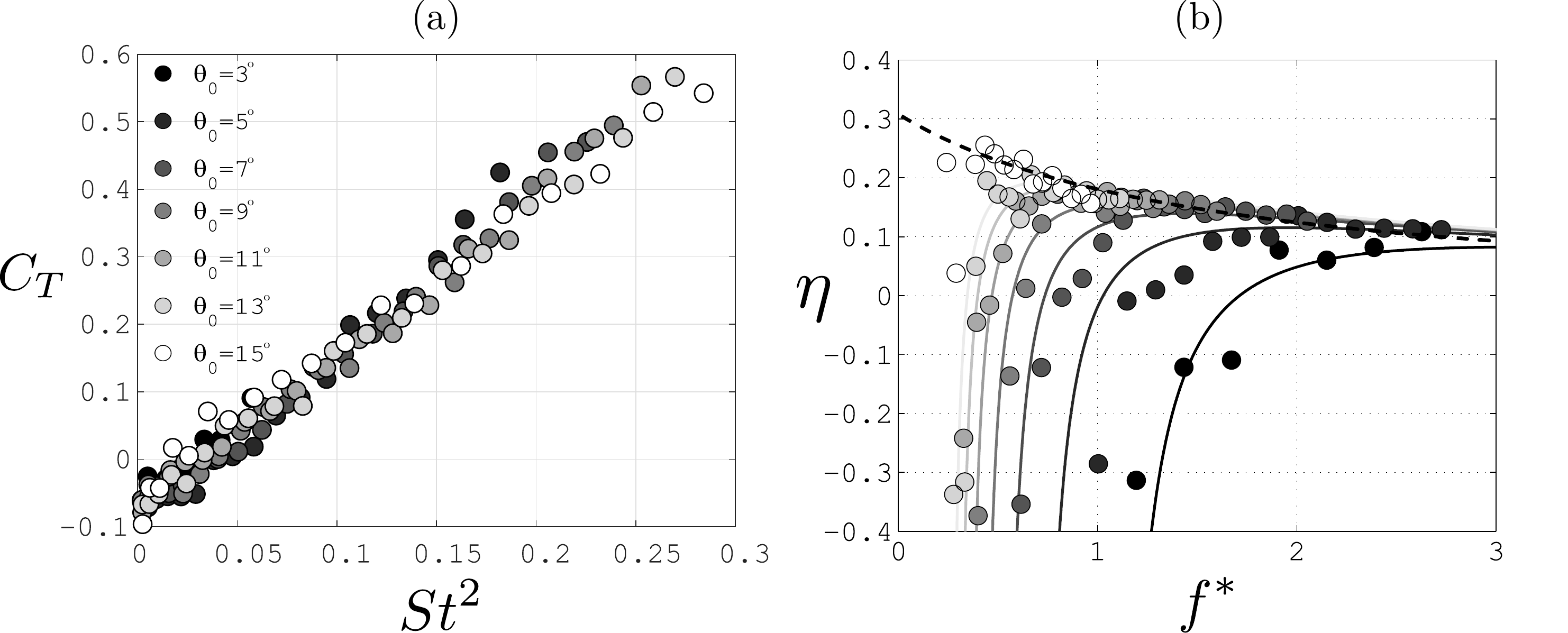}
\caption{Time-averaged (a) thrust and (b) efficiency for a pitching\index{pitch} two-dimensional teardrop foil. Lines represent the the analytical solutions, where the solid lines include viscous drag\index{drag!viscous drag}, and the dashed line is the inviscid approximation. Taken from \cite{Floryan2016}.}
\label{fig:pitchPerf}
\end{figure}

In much of the current literature, the thrust, power, and efficiency\index{efficiency!swimming} are shown as a function of Strouhal number\index{Strouhal number} to define the performance envelope (Reynolds number\index{Reynolds number} effects are often neglected, but more about that later). Figure \ref{fig:pitchPerf} shows the thrust and efficiency of a two-dimensional pitching\index{pitch} propulsor for a range of frequencies and amplitudes. The thrust goes as $C_T\sim St^2$ for all of the pitch amplitudes, which is in accord with equation \ref{pitch_scale}, originally proposed by Floryan et al. \cite{Floryan2016}.  They also showed that for inviscid flow the efficiency\index{efficiency!swimming} should follow a decaying curve with $f^*$, which is in accord with experiments for large values of $f^*$ (see figure~\ref{fig:pitchPerf}b).  At lower frequencies, the efficiency decreases dramatically due to the viscous drag\index{drag!viscous drag}, which drives the thrust\textemdash and therefore the efficiency\textemdash negative. 

\begin{figure}
\centering
\includegraphics[width=1\textwidth]{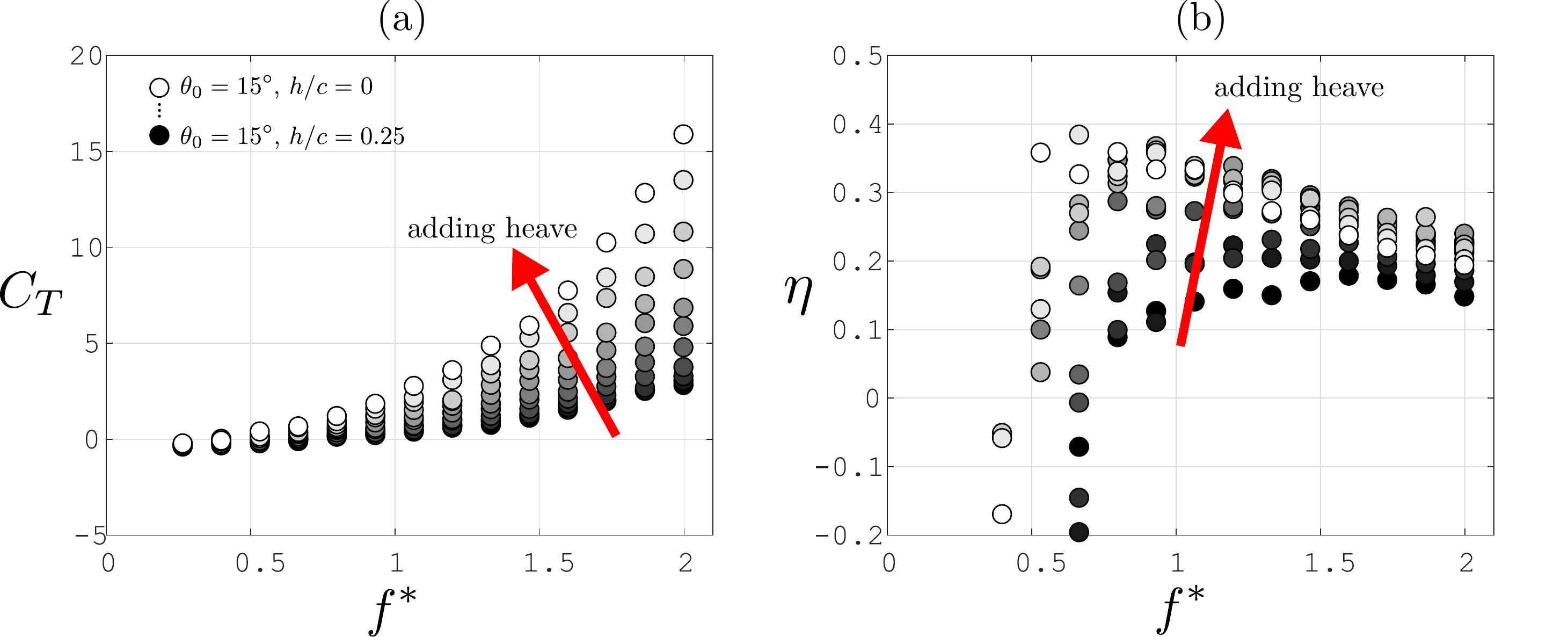}
\caption{Time-averaged (a) thrust and (b) efficiency for a two-dimensional teardrop foil with incremental increases in heave\index{heave} amplitude to a pitching\index{pitch} foil. Phase between pitch and heave is $\phi=270^\circ$. Taken from \cite{VanBuren2018_1}.\index{thrust}}
\label{fig:pitchHeave}
\end{figure}

On their own, pitching\index{pitch} and heaving\index{heave} motions are not very fish-like in appearance, and they only reach 20-30\% peak propulsive efficiency\index{efficiency!swimming}. When these motions are combined, however, they can resemble fish-like motions much more closely, and achieve significant improvements in performance. Figure \ref{fig:pitchHeave} exemplifies how simply adding heave\index{heave} to a pitching\index{pitch} motion can dramatically increase the thrust and efficiency\index{efficiency!swimming} of the propulsor. The phase difference $\phi$ between the pitch\index{pitch} and heave\index{heave} motions governs the motion path of the foil, and to be most fish-like the trailing edge of the foil must lag the leading edge such that the foil slices through the water.  This reduces the maximum angle of attack\index{angle of attack} of the foil, and the likelihood of flow separation\index{flow separation}.  Many researchers suggest that $\phi=270^\circ$ maximizes the propulsive efficiency \cite{quinn2015, xu2016, VanBuren2018_1}. The scaling arguments for pitching\index{pitch} or heaving\index{heave} motions presented by Floryan et al. \cite{Floryan2016} have since been successfully extended to simultaneously pitching\index{pitch} and heaving\index{heave} motions in Van Buren et al. \cite{VanBuren2018_1}.

There are other ways to manipulate the propulsor performance. For instance, Van Buren et al. \cite{VanBuren2017} showed that the thrust production is strongly correlated to the trailing edge velocity of the propulsor in both pitch\index{pitch} and heave\index{heave}.  For a sinusoidal motion\index{motion!sinusoidal}, the trailing edge velocity can be changed by adjusting the frequency or amplitude of motion.  However, a motion profile that is more like a square wave\index{motion!non-sinusoidal} can have a much higher peak trailing edge velocity than a sinusoidal motion of equal amplitude and frequency. By making the motion more square-like, the thrust production could be increased by over 300\%. 

If energy expenditure is more important than thrust, you might consider intermittent motions\index{motion!intermittent}. Floryan et al. \cite{Floryan2017} showed that by swimming intermittently, the energy expenditure decreased linearly with duty cycle when compared to continuous swimming, although the average swimming speed\index{swimming speed} is reduced accordingly.   In general, other factors such as the metabolic rate\index{metabolic rate} will need to be taken into account, and these will generally reduce the benefits of intermittent swimming.

Lastly, one can achieve a significant performance boost by swimming near a boundary, or in the wake\index{wake} of another swimmer. Quinn et al. \cite{quinn2014_2, quinn2014_3} showed experimentally that when a propulsor swims near a solid boundary it experiences a thrust increase due to ground effect\index{ground effect}.  Similarly, two propulsors can significantly improve performance by swimming side-by-side \cite{dewey2014propulsive}. The maximum thrust is found for a phase difference of 180$^\circ$, and the maximum efficiency\index{efficiency!swimming} is found for 0$^\circ$. In both instances, two paired foils could produce higher thrust and efficiency than a single foil. Experiments and simulations have also demonstrated the advantages of fish schooling, indicating that one swimmer can benefit from being in the wake of the other \cite{boschitsch2014propulsive,maertens2017}.

\subsubsection{\emph{Flexibility}}\index{flexibility}

So far, we have considered only rigid propulsors, but most fish exhibit some form of passive/active flexibility\index{flexibility} in their propulsive fins while swimming.  This flexibility turns out to be a valuable asset in increasing the propulsive performance. For example, Quinn and Dewey \cite{dewey2013,dewey2013thesis, quinn2015thesis,quinn2015} explored the influence of adding flexibility to simple pitching\index{pitch} and heaving\index{heave} propulsors. They found that flexibility adds a hierarchy of resonances\index{resonance} to the system, which can be modeled using the linear beam equation. That is,  
\begin{equation}
\rho s c \frac{\partial^2e}{\partial t^2}+EI \frac{\partial^4e}{\partial x^4}=F_{ext},
\end{equation}
where $e$ is a small panel deflection, $EI$ is the panel stiffness, and $F_{ext}$ is the force from the fluid. In the usual beam equation, the first term uses properties of the beam, but here it is expected that the virtual mass forces due to the water surrounding the panel will dominate (panel mass $\ll$ virtual mass), and so fluid properties are used instead (see also \cite{allen2001energy}).  This yields a new non-dimensional frequency 
\begin{equation}
\hat{f}^*=f(\rho s c^5/(EI))^{1/2},
\end{equation}
which is the ratio of the frequency of the driving motion to the first resonant frequency of the panel when added mass\index{added mass} is much larger than panel mass.

\begin{figure}
\centering
\includegraphics[width=1\textwidth]{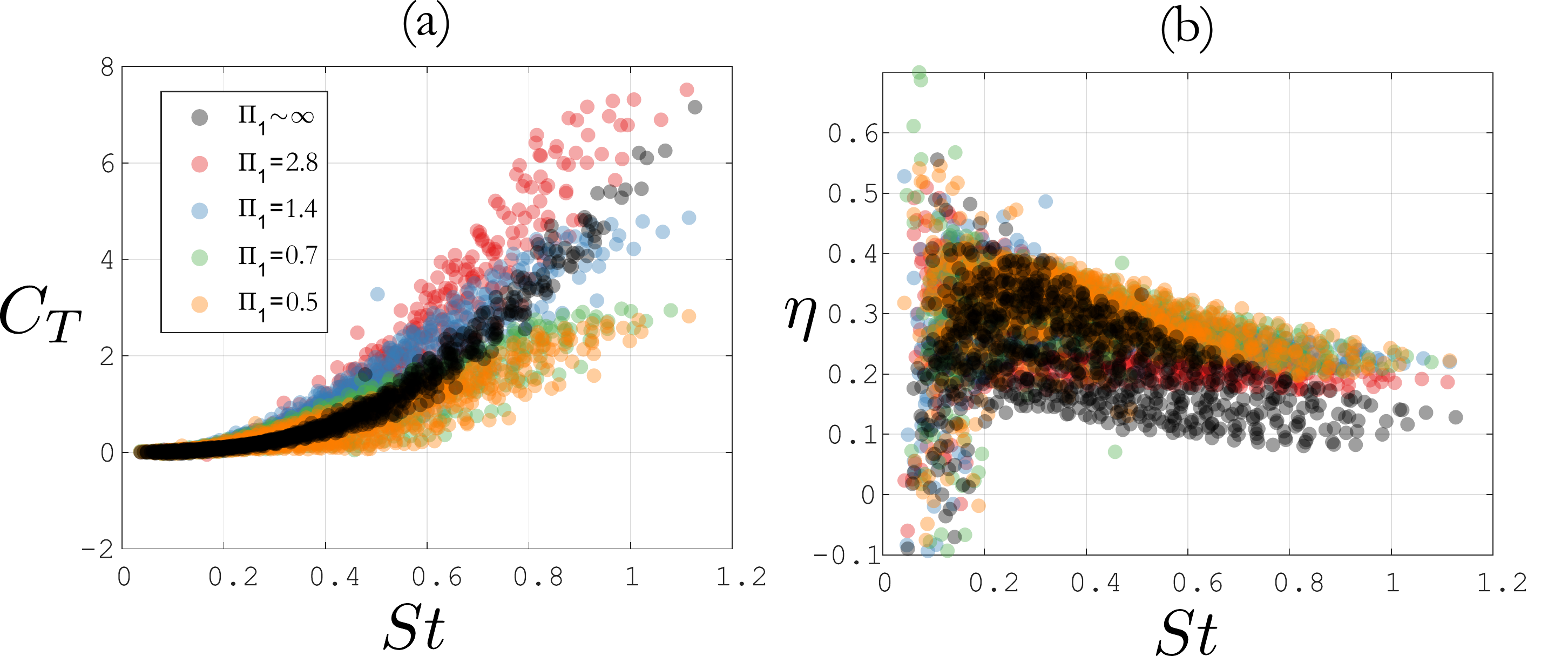}
\caption{Effects of flexibility\index{flexibility} on time-averaged (a) thrust and (b) efficiency for a pitching\index{pitch} and heaving\index{heave} panels separated by phase $\phi=270^\circ$. Pitch amplitudes $\theta=\{6^\circ$, 9$^\circ$,...15$^\circ$\}; heave amplitudes $h_0$/$c=\{0.083$, 0.167,...0.33\}; and frequencies $f=\{0.2$, 0.25,...1 Hz\}. The rigid panel corresponds to $\Pi_1 \sim \infty$.  }
\label{fig:flexPerf}
\end{figure}

Let us consider a simple panel in fish-like motion (pitching\index{pitch} and heaving\index{heave} separated by $\phi=270^\circ$) of varying flexibility\index{flexibility} ($\Pi_1$) and compare that to a stiff panel ($\Pi_1 \sim \infty$), where 
$$\Pi_1=\frac{Et^3}{12(1-\nu_p^2)\rho U_\infty^2 c^3}.$$
Here $\nu_p$ is the Young's modulus of the panel, and $t$ is the panel thickness.  The thrust and efficiency\index{efficiency!swimming} of these panels, covering the range of flexural stiffnesses seen in biological propulsors, are shown in figure \ref{fig:flexPerf} for a large parameter space of leading edge motions.  For certain flexibilities, there is a clear envelope where the thrust production is up to 1.5 times higher than the thrust of a rigid panel. This is because the frequency of the panel motion is close to the resonance\index{resonance} of the system, and the trailing edge amplitude becomes amplified.  The peak efficiency is less influenced by flexibility\index{flexibility} than thrust, but the efficiency tends to remain higher over a larger Strouhal number\index{Strouhal number} range.

\begin{figure}
\centering
\includegraphics[width=1\textwidth]{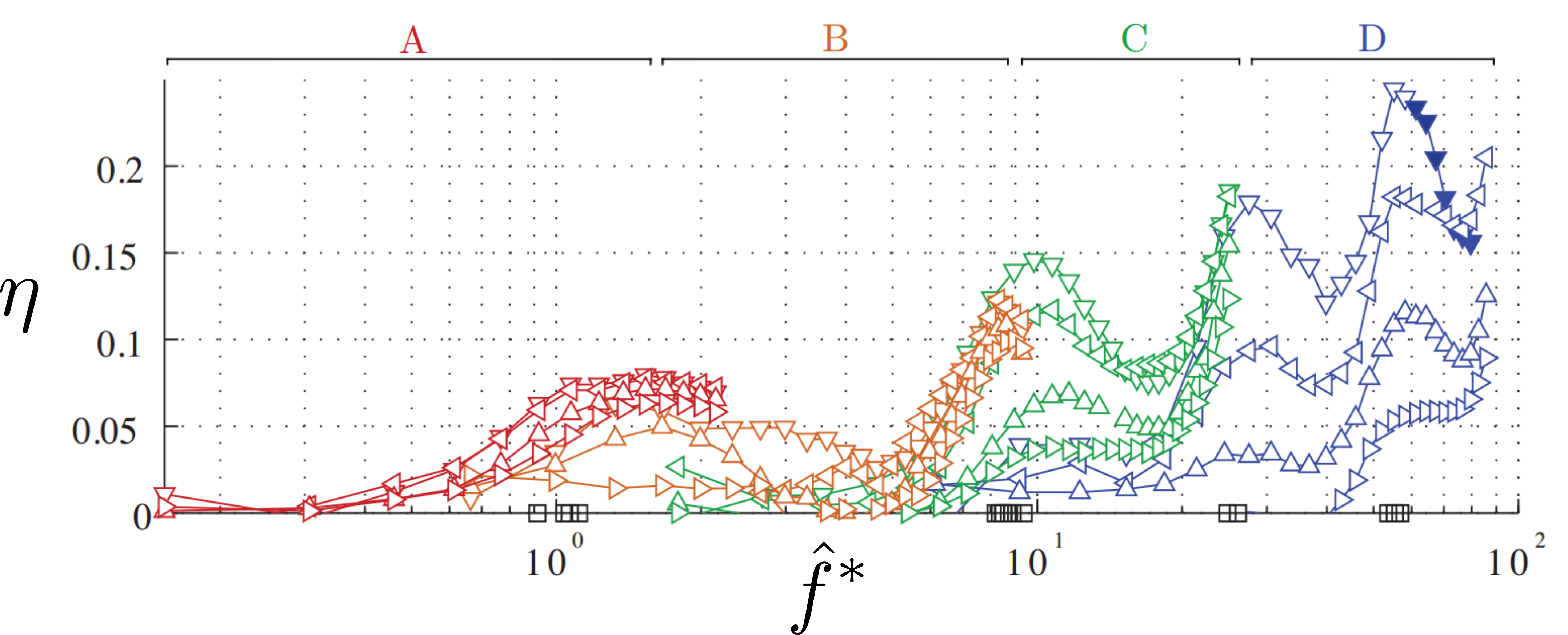}
\caption{Efficiency peaks of heaving\index{heave} flexible panels at multiple resonance\index{resonance} modes. Panels A,B,C,and D have stiffnesses $EI=3.2\times10^{-1}$, 1.1$\times10^{-2}$, 8.1$\times10^{-4}$, 6.9$\times10^{-5}$, and are colored red, orange, green, and blue respectively. Taken from \cite{quinn2014}.}
\label{fig:flexEff}
\end{figure}

The work of Dewey et al.\cite{dewey2013} found similar trends to these for pitching\index{pitch} panels of varying aspect ratio, over the same range of flexibilities.  Quinn et al. \cite{quinn2014} extended this work to very flexible panels in heave\index{heave}, where more than one resonance\index{resonance} mode could be excited (similar to undulatory swimmers).  They found that efficiency\index{efficiency!swimming} peaks occurred close to each resonant frequency, as shown in figure \ref{fig:flexEff}, indicating that a simple analysis based on linear beam theory can sometimes be a valuable tool in predicting the performance of flexible propulsors.

\subsubsection{\emph{Concept design}}\index{underwater vehicle}\index{concept design}

\begin{figure}
\centering
\includegraphics[width=0.65\textwidth]{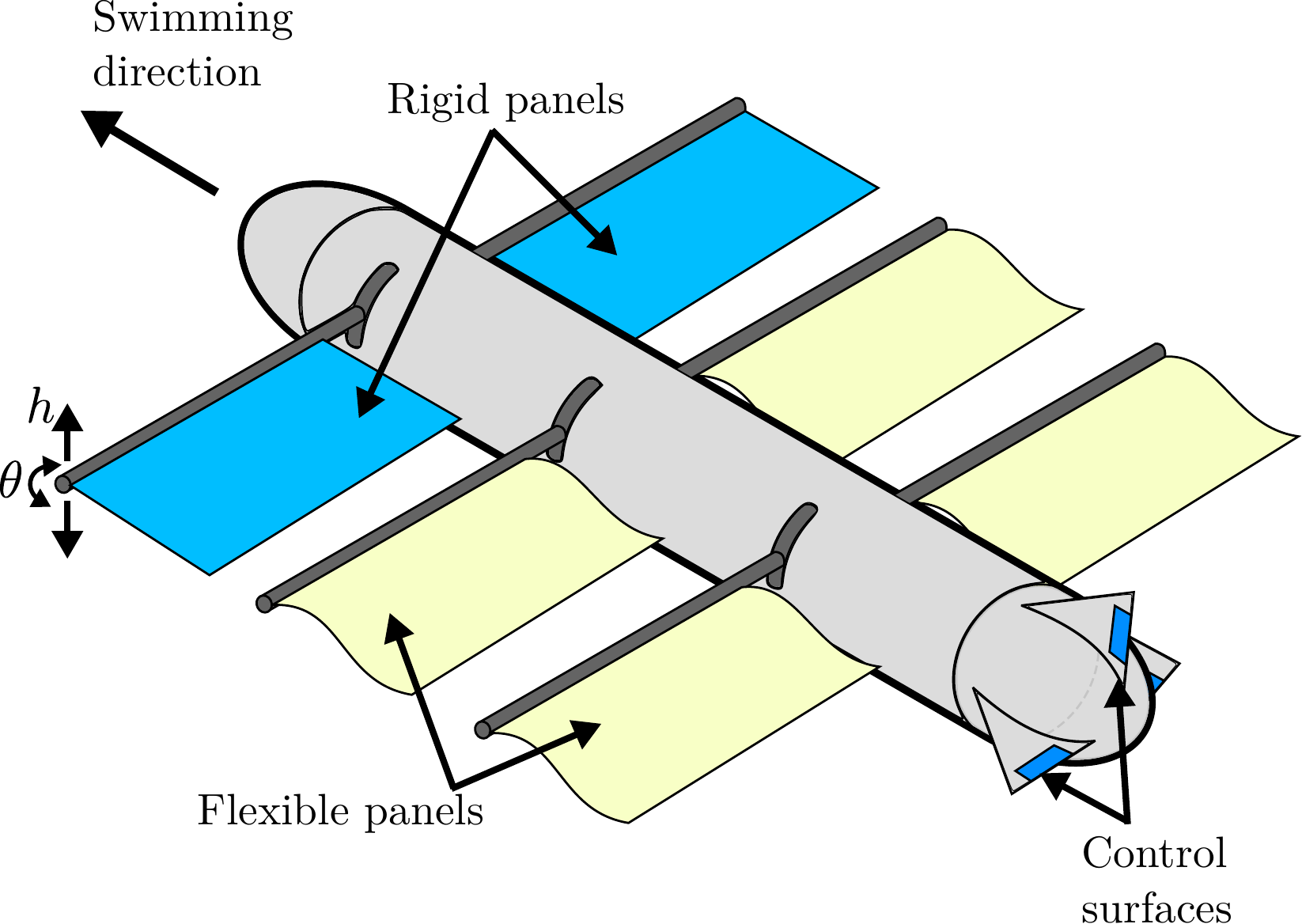}
\caption{Concept design\index{concept design} that uses undulatory and oscillatory propulsion, taking advantage of fin-fin interaction\index{fin-fin interaction} and flexibility\index{flexibility} performance benefits.}
\label{fig:concept1}
\end{figure}

Consider the concept design\index{concept design} using oscillatory and undulatory motions shown in figure \ref{fig:concept1}. The long tube-shaped body is similar to many underwater vehicles in use today, with control surfaces for stability and maneuvering located near the stern.

The propulsors are a series of rigid and flexible pitching\index{pitch} and heaving\index{heave} panels whose motions can be individually controlled. The first pair of propulsors near the bow are rigid, which can be used to provide thrust through lift-\index{lift} and added mass-\index{added mass}based forces, and maneuvering through drag-based forces\index{drag}. The other pairs of propulsors are flexible to maximize thrust and efficiency\index{efficiency!swimming} of swimming. Multiple propulsors are intended to exploit fin-fin interactions\index{fin-fin interaction}. The motion profiles of all pairs of propulsors can be individually tuned and may take advantage of non-sinusoidal\index{motion!non-sinusoidal} motions to achieve the thrust benefits associated with higher trailing edge velocities.

This type of vehicle is likely to be used where high efficiency and quiet swimming are valued.  For long missions, the surface of the rigid panels could house flexible solar panels, so that the vehicle can occasionally recharge its batteries at the water surface, making it fully independent. 

\subsection{Pulsatile jet}\index{swimming type!pulsatile}
Pulsatile jet swimmers include squid\index{swimming animals!squid}, jellyfish\index{swimming animals!jellyfish}, and mollusks\index{swimming animals!mollusk}.  They propel themselves forward by impulsively injecting momentum into the surrounding fluid, producing forward thrust. Pulsatile jet swimmers have been widely studied in relation to jellyfish\index{swimming animals!jellyfish} (see, for example, \cite{dabiri2010}), but our general understanding is enriched by research on starting jets, vortex\index{vortex!ring} ring evolution, and synthetic jets (a flow control device that creates a train of vortex\index{vortex!ring} rings \cite{GlezerAmitay2002}).

\begin{figure}
\centering
\includegraphics[width=0.85\textwidth]{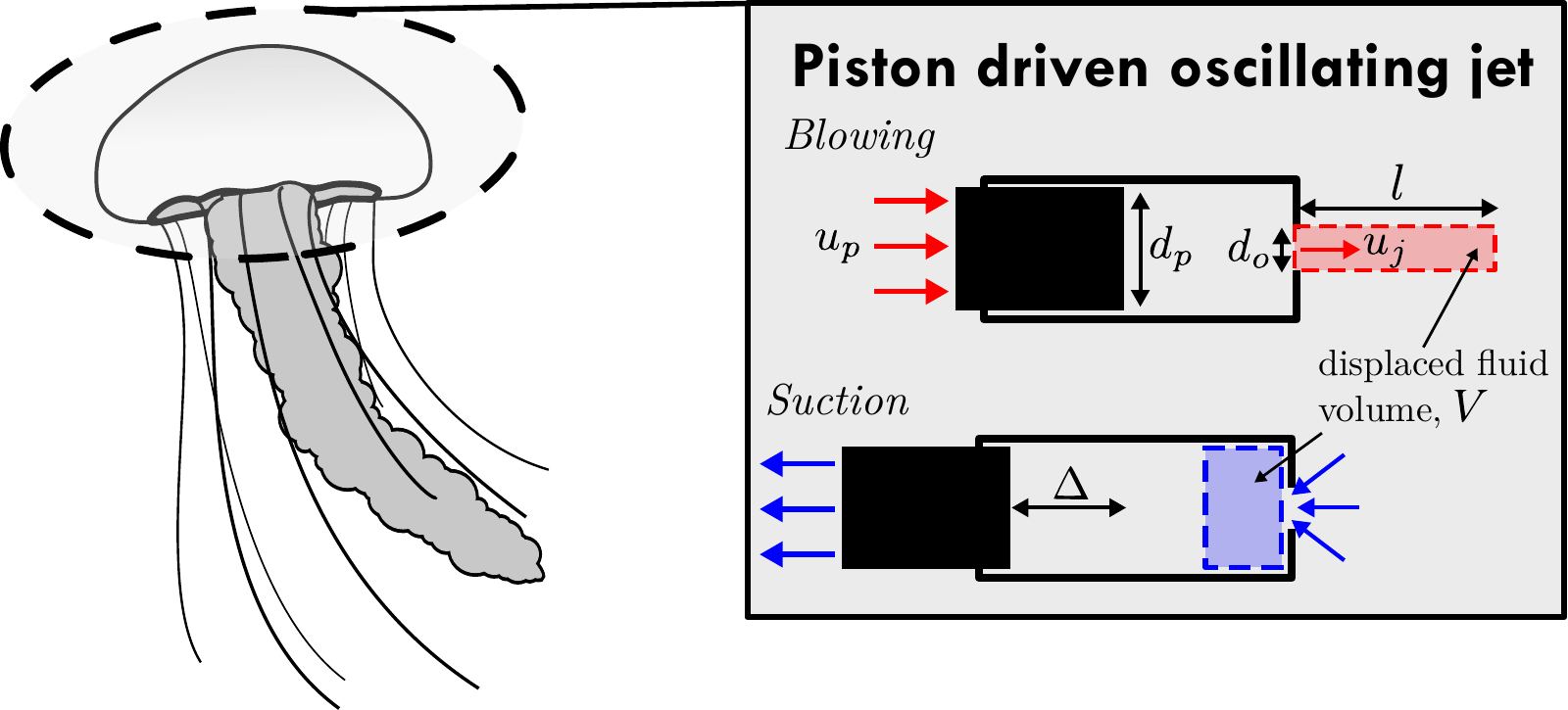}
\caption{Example of how one could simplify a pulsatile swimmer into a piston driven oscillating jet\index{jet}.}
\label{fig:propulsorSimple2}
\end{figure}

We can model a pulsatile jet propulsor as a driver (e.g., a piston or a piezoelectric membrane) that displaces a fluid within a cavity that has a small opening to produces a jet. If the fluid driver oscillates periodically, the jet becomes pulsatile.  This type of simplification is shown schematically in figure \ref{fig:propulsorSimple2}, where a piston with diameter $d_p$ oscillates at peak velocity $u_p(t)$ and peak amplitude $\Delta$, displacing a fluid volume $V$ through an orifice with diameter $d_o$ creating a time-varying jet with exit velocity $u_j(t)$. 

\subsubsection{\emph{Wake characteristics}}\index{wake}

\begin{figure}
\centering
\includegraphics[width=0.75\textwidth]{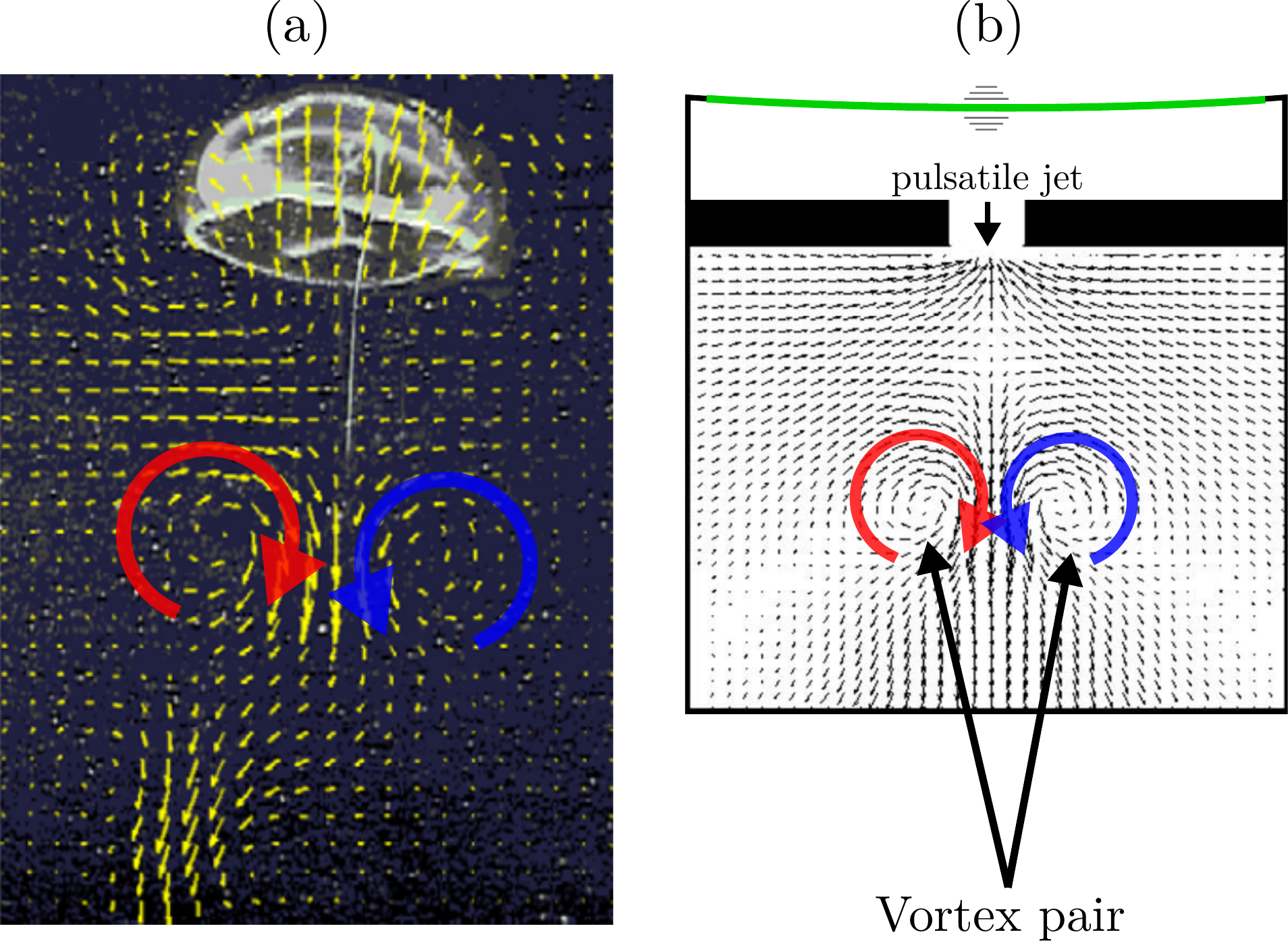}
\caption{Velocity field measurement of a (a) jellyfish\index{swimming animals!jellyfish} and (b) synthetic jet\index{jet}. Both show similar vortex\index{vortex!pair} pairs generated at the orifice and propagating away at their own induced velocity. Adapted from \cite{dabiri2010, VanBurenetal2013_2}\index{flow visualization}.}
\label{fig:jellyfishWake}
\end{figure}

As we did in the previous section, we can examine the wake to give us insight into the thrust mechanisms of pulsatile swimmers. Dabiri et al. \cite{dabiri2010} directly measured the velocity field downstream of jellyfish\index{swimming animals!jellyfish} and showed that their wakes are primarily made up of a train of vortex\index{vortex!ring} rings produced by the periodic expulsion of water.  Figure \ref{fig:jellyfishWake} shows a sample flow measurement of a jellyfish\index{swimming animals!jellyfish} wake (three-dimensional) and compares it to a snapshot of the flow produced by a synthetic jet\index{jet} (two-dimensional).  The wakes are qualitatively similar in that they create successive vortex\index{vortex!ring} rings or vortex\index{vortex!pair} pairs that convect away from the orifice at their own induced velocity.  The vortices generate a region of high velocity on the centerline that represents the bulk of the momentum that is injected into the fluid to produce thrust.

The vortex\index{vortex!ring} ring generated by a circular orifice (like a jellyfish\index{swimming animals!jellyfish}) is stable and simply diffuses as it evolves downstream.   A vortex\index{vortex} loop generated by a non-axisymmetric orifice (like a scallop or clam) is unstable, and it can exhibit periodic axis-switching downstream until it reaches an axisymmetric equilibrium \cite{DhanakBernardinis1981, VanBurenetal2013_1}. This instability can induce vortex\index{vortex} pinch-off and an earlier breakdown, which can affect the thrust production.

\subsubsection{\emph{Vortex formation and evolution}}\index{vortex}

\begin{figure}
\centering
\includegraphics[width=0.5\textwidth]{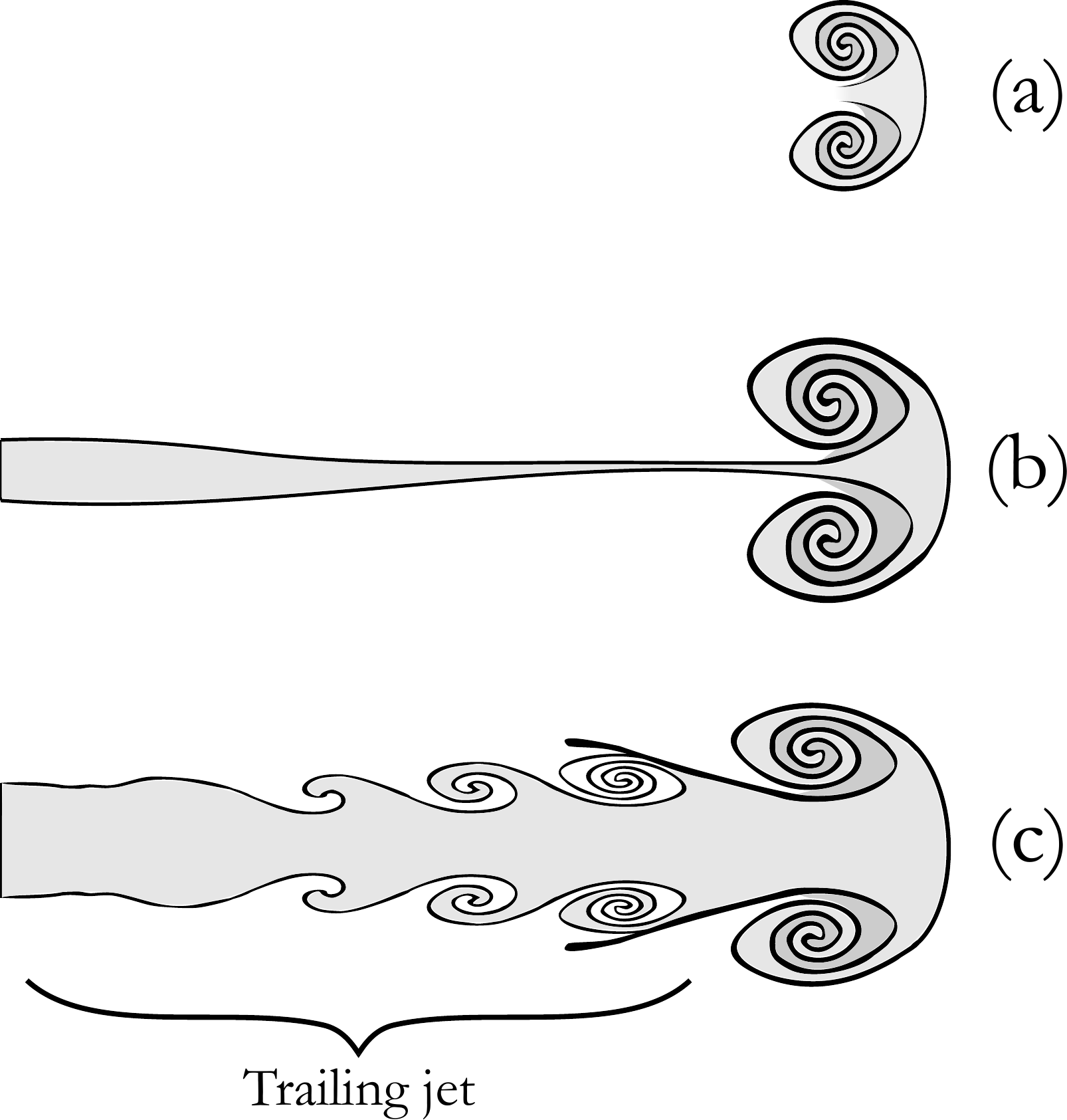}
\caption{Illustration of the vortex\index{vortex} formation from a piston driven cavity with a circular orifice at stroke ratios\index{stroke ratio} (a) $l/d_o=2$, (b) 3.8, and (c) 14.5. After the critical stroke ratio of $l/d_o=4$, the appearance of a trailing jet\index{jet} is observed. Vortex structure inspired from \cite{gharib1998}.}
\label{fig:vortexFormation}
\end{figure}

A vortex\index{vortex!ring} ring is formed when a volume of fluid is driven impulsively through a circular orifice in the cavity, and the boundary layers\index{boundary layer} on the walls of the cavity separate at the orifice edges resulting in a growing roll-up of vorticity. The volume of fluid that exits the orifice can be estimated as a column with length $l$ and diameter $d_o$. The aspect ratio of this fluid column is called the stroke ratio\index{stroke ratio}, $l/d_o$, and  Gharib et al.\cite{gharib1998} showed that there is a critical stroke ratio\index{stroke ratio}, $l/d_o \sim 4$, after which the circulation in the main vortex\index{vortex} stops growing and a trailing jet\index{jet} forms behind the vortex. This occurs when the fluid driver cannot provide energy fast enough to continue the vortex growth. Figure \ref{fig:vortexFormation} shows a sketch\index{flow visualization} of a vortex\index{vortex} formed using $l/d_o=2$, 3.8, and 14.5 exhibiting the trailing jet\index{jet} that forms after the critical stroke ratio\index{stroke ratio} is reached.

\begin{figure}
\centering
\includegraphics[width=1\textwidth]{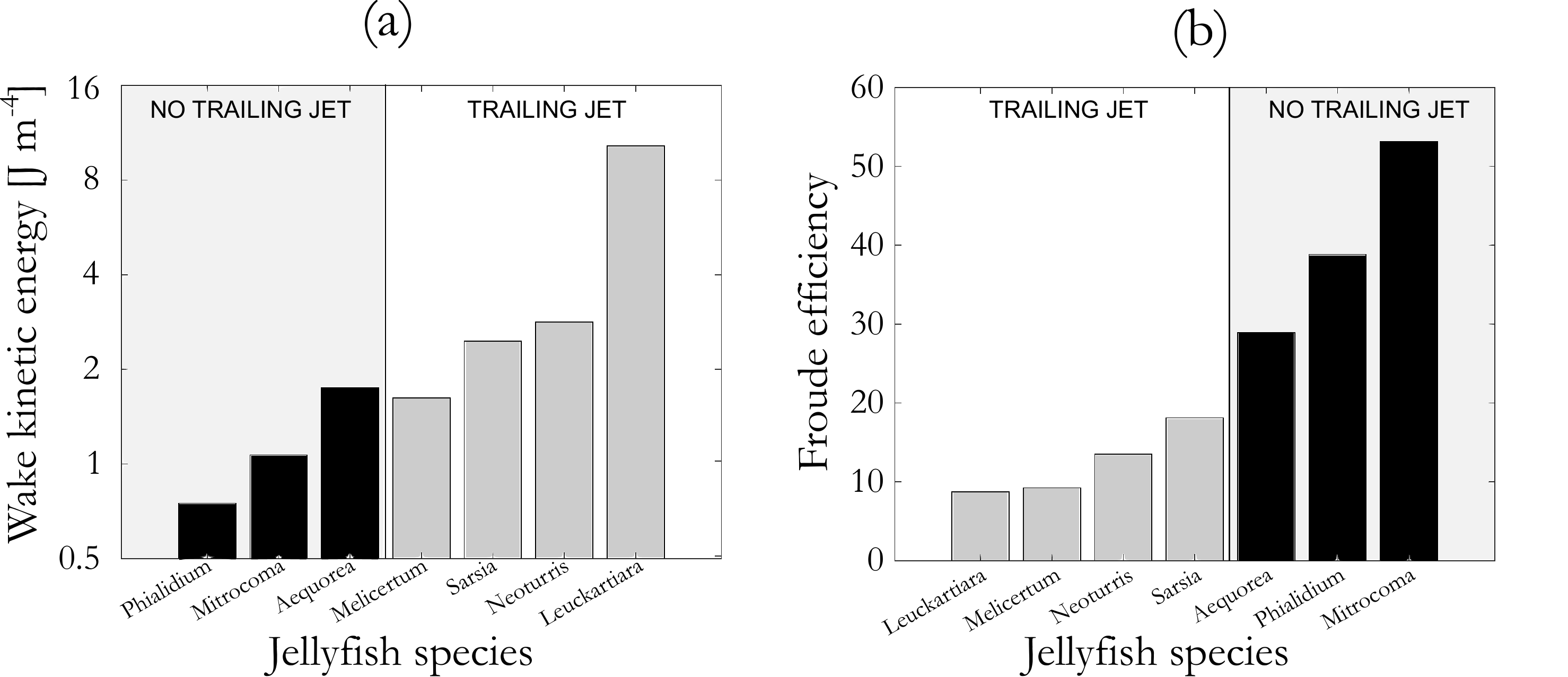}
\caption{Various species of jellyfish\index{swimming animals!jellyfish} (a) wake kinetic energy  and (b) propulsive efficiency while swimming. Four jellyfish commonly exhibit a vortex\index{vortex} with a trailing jet\index{jet}, resulting in higher thrust but much lower efficiency. Data taken from \cite{dabiri2010}.}
\label{fig:jellyfishPerf}
\end{figure}

The trailing jet\index{jet} formation can dramatically influence swimming performance in jellyfish\index{swimming animals!jellyfish}. Dabiri et al. \cite{dabiri2010} studied seven jellyfish species of various sizes, and tracked their vortex\index{vortex} formation while swimming.  Three of the jellyfish species produced vortices below the critical stroke ratio\index{stroke ratio}, while the other four exceeded the critical stroke ratio so that they produced vortex\index{vortex!ring} rings with a trailing jet\index{jet}.  Figure \ref{fig:jellyfishPerf} shows that the jellyfish that commonly exceeded the critical stroke ratio swam faster than their counterparts, but much less efficiently. From a design perspective, this means that by changing the stroke ratio\index{stroke ratio} one could either favor thrust production or efficiency\index{efficiency!swimming}.

Now, for a zero-net-mass-flux condition, any vortex\index{vortex} created through exhalation needs an equal and opposite inhalation (or suction) to refill the cavity.  For example, figure \ref{fig:jellyfishWake}b shows the moment of the suction phase of a synthetic jet\index{jet} where fluid is inhaled into the cavity. The suction comes from all directions, and with relatively low peak velocities.  Hence the negative thrust produced during the suction phase is relatively small compared to that seen in the blowing phase which directs all of the momentum in one direction.

In addition, the vortex\index{vortex} spacing in a series of pulses can strongly impact the resulting velocity field. For synthetic jets\index{jet}, the Strouhal number\index{Strouhal number} is defined as
\begin{equation}
St=\frac{f_{j}\,d_o}{U_j}
\end{equation}
where $f_{j}$ is the jet\index{jet} frequency and $U_j$ is some characteristic velocity of the jet\index{jet}, either the time-average of the absolute streamwise velocity or the peak velocity. This is effectively the inverse of the stroke ratio\index{stroke ratio} if we recognize that $U_j/f_{j} \sim l$. Holman et al. \cite{Holmanetal2005} showed that there is a minimum stroke ratio that must be achieved to create a jet\index{jet}. If the stroke ratio is too small, the vortex\index{vortex!pair} pairs become too close together, and the vortices created by one blowing cycle could be pulled back into the orifice by the next suction cycle.  For two-dimensional flows, the minimum stroke ratio is $l/d_o \approx 1$ (the actual number depends on orifice geometry). Thus, we can define a range of stroke ratios, $1 \geq l/d_o \leq 4$, for which a train of trailing jet-free vortices can be formed.

\subsubsection{\emph{Thrust production}}
\index{thrust}Consider the force required to generate a vortex\index{vortex} with circulation $\Gamma=\oint {\bf v} \cdot d{\bf S}$ where ${\bf v}$ is the fluid velocity on closed surface ${\bf S}$.  The circulation can be thought of as a measure of the rotation rate of a vortex\index{vortex}, and we can relate it to the equivalent impulse\index{impulse} (integral of force over time) according to \cite{huggins1966}
\begin{equation}
I=\int F_x dt = \frac{1}{2}\pi d_o^2 \, \rho \, \Gamma .
\end{equation}
Consequently, the mean thrust force per pulse is $\overline{F_x}=I/t_p$ where $t_p$ is the time of a single pulse.  

The circulation is an output of the system, and it would be helpful to be able to predict the force based upon known inputs. We can start by using a simple slug model \cite{shusser2002, dabiri2004} to relate the time-change in circulation with the jet\index{jet} velocity.  That is,
\begin{equation}
\frac{d\Gamma}{dt}\approx\frac{1}{2}u_j^2.
\end{equation}
A correction to $u_j$ needs to be applied to account for the growth of the boundary layer\index{boundary layer} within the orifice, which is especially important for high stroke ratios\index{stroke ratio}.  Hence,
\begin{equation}
u_j^*=u_j \Bigg( 1+  8\sqrt{\frac{l/d_o}{\pi Re}} \, \Bigg),
\end{equation}
where $Re$ is the Reynolds number\index{Reynolds number} based on $U_j$ and $ d_o$. A simple conservation of mass relation can relate the input piston velocity to the jet\index{jet} velocity, where
\begin{equation}
u_j = u_p \bigg( \frac{d_p}{d_o} \bigg) ^2
\end{equation}
for a circular piston and orifice. Using these simplified relations, the thrust generated by a series of vortex\index{vortex!ring} rings can be estimated. These relations stem from connecting the circulation estimates presented by various researchers \cite{shusser2002, dabiri2004} to the force that would be required to create that circulation \cite{huggins1966}. They can be used to understand how the thrust force might vary with input parameters, although, to our knowledge, they have not yet been verified against experiment.

\begin{figure}
\centering
\includegraphics[width=1\textwidth]{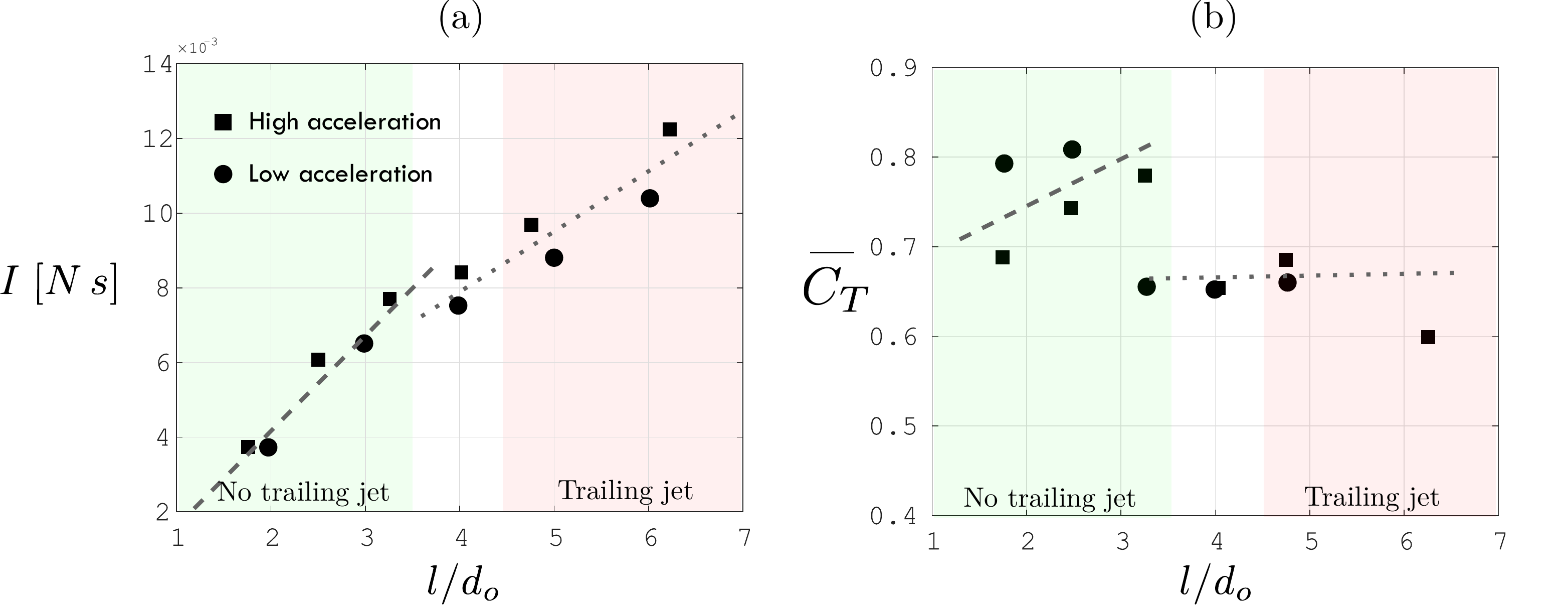}
\caption{Impact of stroke ratio\index{stroke ratio} on the (a) impulse\index{impulse} and (b) thrust coefficient\index{coefficient!thrust}. High and low accelerations considered for vortex\index{vortex!pair} pairs with and without trailing jets\index{jet}. Data taken from \cite{krueger2003} with \emph{rough} trend lines added.}
\label{fig:impulse}
\end{figure}

Krueger and Gharib \cite{krueger2003} directly measured the impulse\index{impulse} and thrust production of a piston-driven vortex\index{vortex} generator, varying the peak acceleration of the piston.  Their results are shown in figure~\ref{fig:impulse}. The thrust coefficient\index{coefficient!thrust} was defined by
\begin{equation}
C_T = \frac{F_x}{\frac{1}{2}\rho U_j^2 \pi (d_o/2)^2} .
\end{equation}
We have added some lines to highlight trends in the data. For high and low accelerations, the impulse\index{impulse} grows linearly with stroke ratio\index{stroke ratio} because for higher stroke ratios the jet vortex\index{vortex} is formed over a longer period of time.  Note the change in offset and slope of the trend lines at the point where a trailing jet\index{jet} is produced ($l/d_o \ge 4$). Similar trends are seen for the average thrust per pulse, where a penalty occurs at the production of a trailing jet\index{jet}. These simple experiments help explain the performance measurements of Dabiri et al. \cite{dabiri2010} on swimming jellyfish\index{swimming animals!jellyfish}. 

Much still needs to be studied on pulsatile jets\index{jet} as a viable propulsion system. This includes further parametric studies on the force generated by vortex\index{vortex!ring} ring generators and defining and measuring the efficiency\index{efficiency!swimming}. We saw that the stroke ratio\index{stroke ratio} is a critical parameter and that by changing the stroke ratio one could either favor thrust production or efficiency.  In future work, compliant materials that make up the cavity and orifice could be considered. Also, elegant and compact designs would be necessary to propel an operating aquatic vehicle. 

\subsubsection{\emph{Concept design}}\index{underwater vehicle}\index{concept design}

\begin{figure}
\centering
\includegraphics[width=0.75\textwidth]{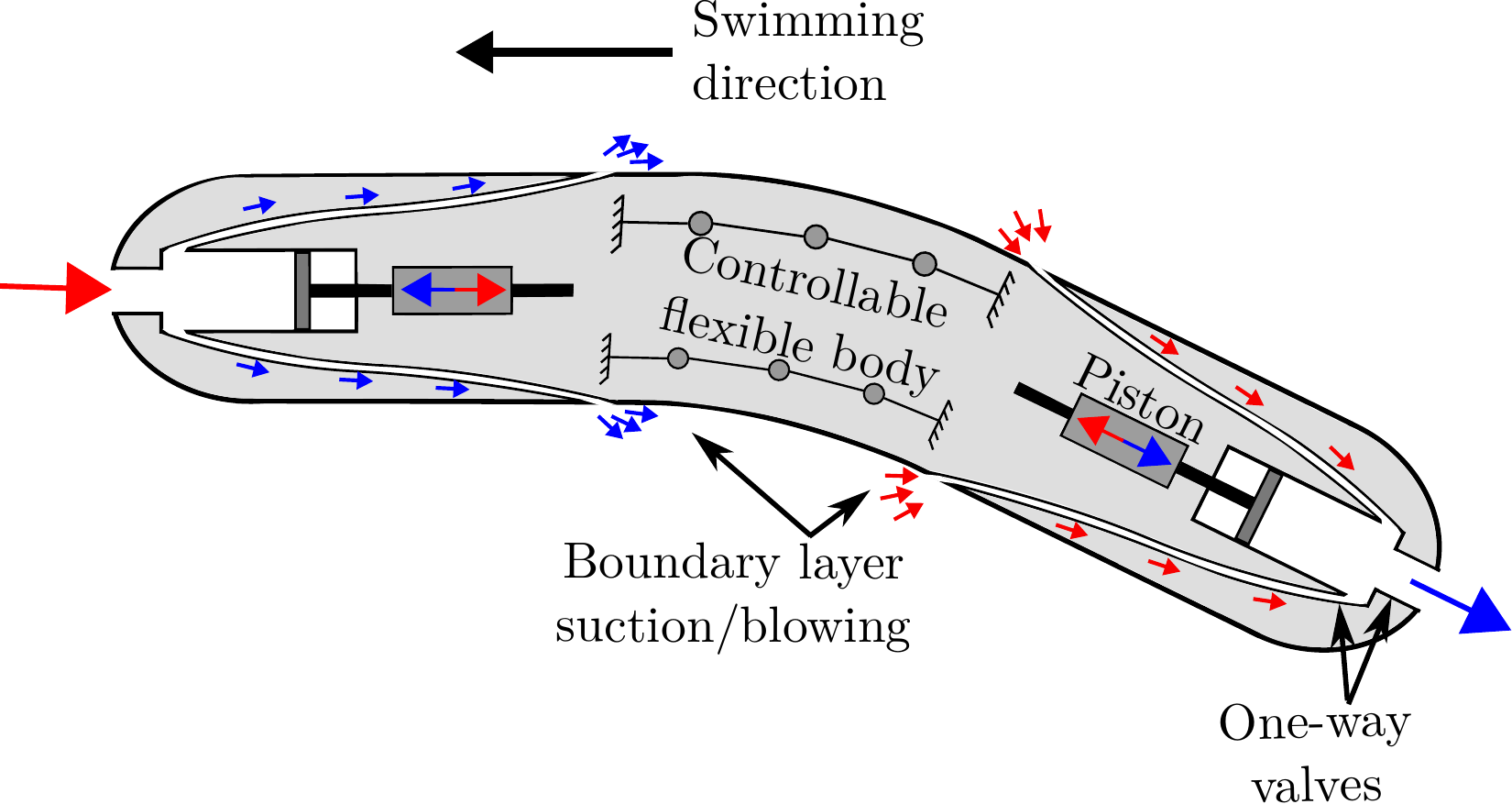}
\caption{Concept design\index{concept design} that uses pulsatile jet\index{jet} propulsion, taking advantage of one-way valves so that both strokes produce thrust, soft robotics to use the body to maneuver, and flow control concepts to limit flow separation\index{flow separation} over the body.}
\label{fig:concept2}
\end{figure}

Figure \ref{fig:concept2} presents a concept design of an aquatic vehicle that uses pulsatile jet\index{jet} propulsion. There are two main oscillating piston-driven cavities, located near the bow and stern. Taking advantage of one-way flow valves, we can ensure that thrust is produced during both halves of the actuation cycle. During the first half-cycle, flow is inhaled at the bow and exhaled at the stern. During the second half-cycle, flow is directed to the surface of the vehicle to either apply blowing or suction within the boundary layer\index{boundary layer}. Through oscillatory blowing and suction on the surface, we can add momentum to the boundary layer\index{boundary layer} to make it less susceptible to separation and mitigate turbulence \cite{Choietal2008, CattafestaSheplak2011}.

To provide steering, we can incorporate a robotic flexible mid-body to maneuver.  As is demonstrated by agile swimmers like the sea lion \cite{fish2003}, excellent maneuvering ability can be achieved from body distortion alone.

This type of design may be capable of high swimming speeds\index{swimming speed}. The torpedo-like body could be made exceptionally streamlined, minimizing drag\index{drag}, thereby helping to increase efficiency\index{efficiency!swimming}.  The thrust and efficiency would be primarily dictated by the design of the pistons,  Also, this design concept could be made relatively small, which may be ideal for remote sensing devices.

\subsection{Drag-based}\index{swimming type!drag-based}

Drag-based propulsion is primarily used by amphibious creatures such as turtles\index{swimming animals!turtle} and ducks\index{swimming animals!duck} who have evolved to be proficient in land and water travel. As we saw in figure \ref{fig:swimmerAbility}, drag-based swimmers are not very efficient but competitive in relative swimming speed\index{swimming speed}. We will show that they still have much to offer as inspiration for propulsion systems. 

\begin{figure}
\centering
\includegraphics[width=0.75\textwidth]{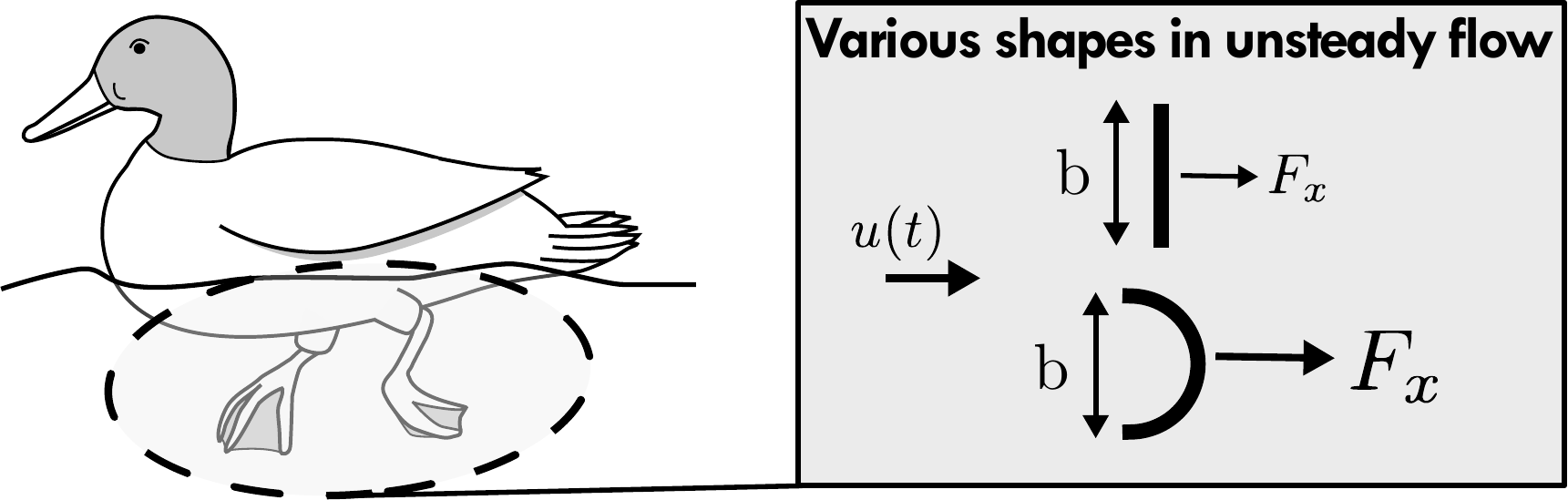}
\caption{Modeling a drag based swimmer for fundamental study. Various shapes have different drag characteristics for similar projected projected lengths $b$.}
\label{fig:propulsorSimple3}
\end{figure}

A simple model for a drag-based propulsor is shown in figure \ref{fig:propulsorSimple3}. Various bluff body\index{bluff body} shapes provide different drag characteristics for similar size and flow velocity, so the planform and aspect ratio are important.  For application to underwater propulsion\index{underwater propulsion}, the bluff body needs to be moved periodically, but for modest reduced frequencies, the forces and flow fields may be estimated from steady flow considerations. 

\subsubsection{\emph{Wake characteristics}}\index{wake}
The wake of a stationary bluff body\index{bluff body} has been a topic of interest in aerodynamics and hydrodynamics for centuries. 

\begin{figure}
\centering
\includegraphics[width=0.7\textwidth]{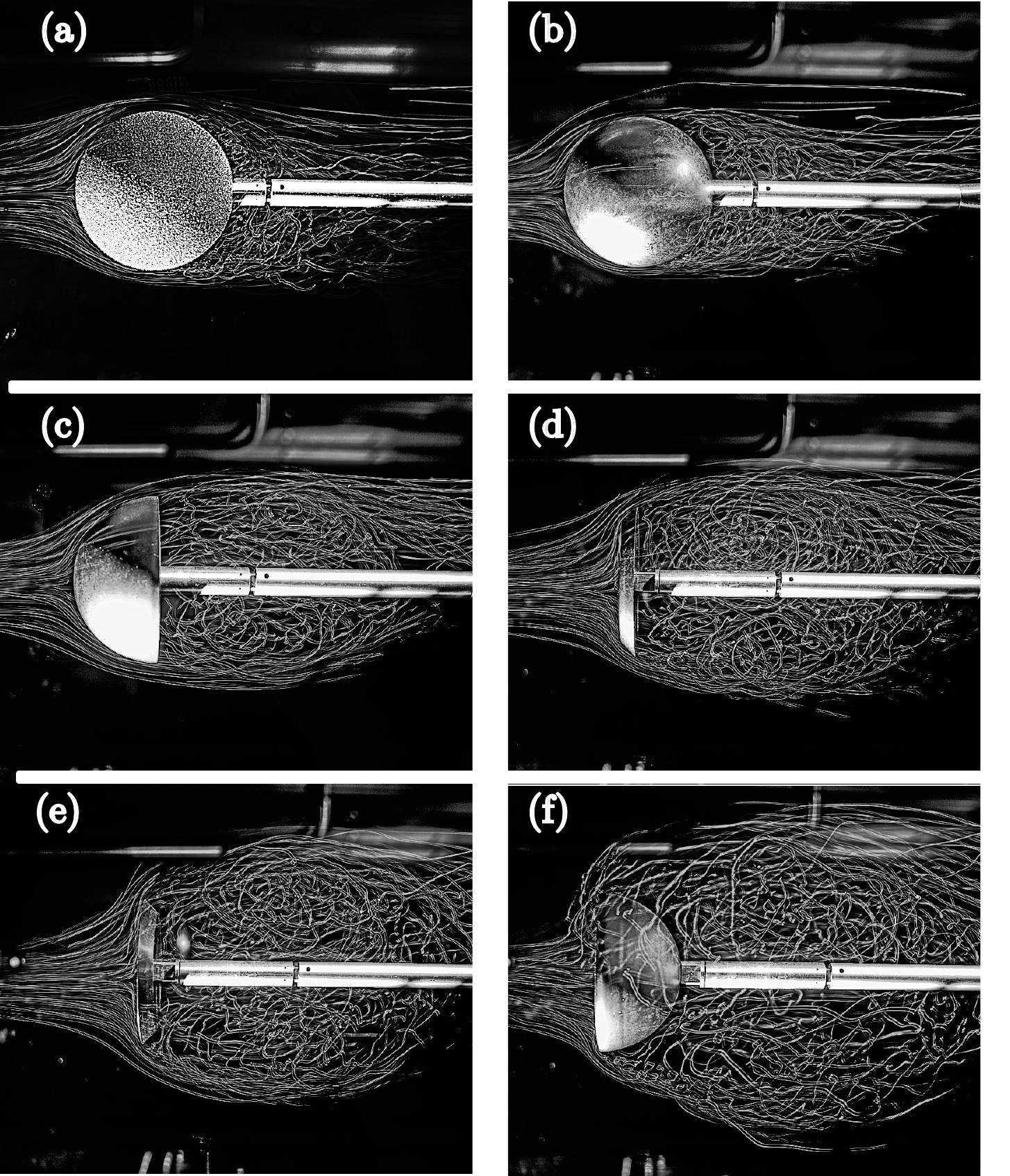}
\caption{Helium bubble flow visualization\index{flow visualization} of air flow around three-dimensional bluff bodies: (a) rough sphere, (b) smooth sphere, (c) forward facing cup, (d) forward facing chamfered plate, (e) backward facing chamfered plate, and (f) reverse facing cup. Images taken in the Rensselaer Polytechnic Institute undergraduate subsonic wind tunnel.}
\label{fig:dragWake}
\end{figure}

As an incoming flow approaches a bluff body\index{bluff body}, it is redirected around it. The fluid near the surface of the body is decelerated due to viscous friction (viscous drag)\index{drag!viscous drag}, and a boundary layer\index{boundary layer} develops. The boundary layer experiences a streamwise pressure gradient which is favorable where the external flow is accelerating, and adverse where it is decelerating. In an adverse pressure gradient, there is a resultant streamwise force acting on the flow inside the boundary layer\index{boundary layer}. If this force is large enough it can force the fluid to actually reverse direction, and the boundary layer\index{boundary layer} is said to separate. Separation is accompanied by a recirculating flow in the wake, and large pressure losses which appear as form drag\index{drag!pressure/form drag}. Form drag due to separation is typically much larger than the viscous drag due to the boundary layer, and so the shape of the body determines the drag since the shape dictates the point of separation and the size of the wake.

This process is illustrated in figure \ref{fig:dragWake}, where the flow around various three-dimensional shapes is visualized using helium bubbles in a wind tunnel. Long exposure photographs reveal the path-lines of the helium bubbles.  We see that in each case the incoming flow first bends around the body, and then at some point separation occurs, and a wake is formed.
The size of the wake dramatically changes for the various shapes, where the spheres have the smallest wake and the reverse facing cup has the largest wake. The drag force generally correlates with wake size, where the larger wake corresponds to the larger drag. The wake can be unstable, like the cylinder exhibited previously in figure \ref{fig:wakes}a, resulting in periodic separation on the body producing a time-varying drag, and a reverse von K\'arm\'an vortex\index{vortex!von K\'{a}rm\'{a}n (reverse)} street. The drag can be directly estimated by measuring the time-averaged momentum deficit in the wake. 

\subsubsection{\emph{Thrust and efficiency}}\index{efficiency!swimming}\index{thrust}
In discussing the thrust and efficiency of drag-based propulsors\index{thrust!drag-based} we will often use the term drag interchangeably with thrust.  Although this may seem confusing, for this class of swimmers the drag produced by the propulsor is, by action-reaction, the thrust of the system. 

As we have indicated earlier, the drag generated by bluff bodies is generally split into viscous and pressure or form drag\index{drag!pressure/form drag}.  For a long thin flat plate oriented parallel to the incoming flow, the major drag component would be viscous drag because the frontal area of the plate is small and there is considerable area for fluid friction to act on the body.  When the plate is perpendicular to the flow, however, there is large form drag, because of the pressure losses due to separation and the formation of a large wake\index{wake}. For drag-based propulsion, we will focus mainly on pressure drag\index{drag!pressure/form drag} because it is easier to produce in large quantity and the flows we are considering have relatively low viscous forces (that is, high Reynolds numbers\index{Reynolds number}).

\begin{table}
\begin{center}
\begin{tabular}{c|c|c|c|c}
Cross-section & Shape (2D) & $C_D$ & Shape (3D) & $C_D$ \\ \hline
\includegraphics[width=0.05\textwidth]{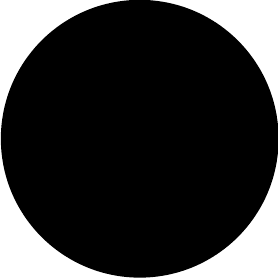} & Cylinder & 1.17 & Sphere & 0.47 \\
\includegraphics[width=0.05\textwidth]{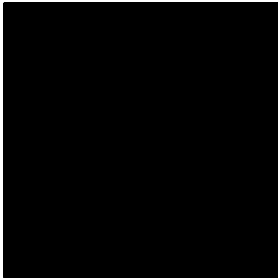} & Square rod & 2.05 & Cube & 1.05 \\
\includegraphics[width=0.05\textwidth]{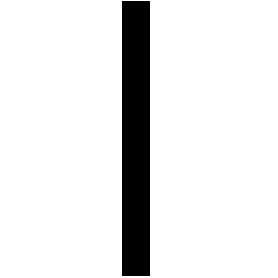} & Flat plate & 1.98 & Circular plate & 1.17 \\
\includegraphics[width=0.05\textwidth]{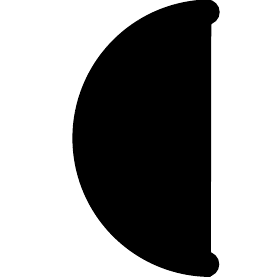} & Hemicylinder (full) & 1.16 & Hemisphere (full) & 0.42 \\
\includegraphics[width=0.05\textwidth]{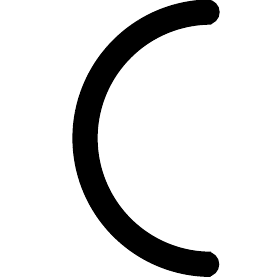} & Hemicylinder (empty) & 1.20 & Hemisphere (empty) & 0.38 \\
\includegraphics[width=0.05\textwidth]{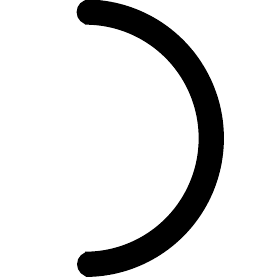} & Hemicylinder (empty) & 2.30 & Hemisphere (empty) & 1.42 \\
\end{tabular}
\caption{Drag coefficients\index{coefficient!drag} for various two- and three-dimensional shapes in a flow going left to right\index{drag}. Data taken from \cite{hoerner1965}.}
\label{tab:Cd}
\end{center}
\end{table}

The drag coefficient\index{drag}\index{coefficient!drag} of a body moving steadily through a fluid is usually defined as
\begin{equation}
C_D = \frac{F_x}{\frac{1}{2}\rho U_p^2sb},
\end{equation}
which is just another form of equation \ref{eqn:drag}, where $U_p$ is the mean velocity of the propulsor with respect to the surrounding fluid (not to be confused with the vehicle velocity, $U_\infty$), $b$ is the projected width of the body, and $s$ is the span-length.  At high Reynolds numbers\index{Reynolds number}, the drag coefficient\index{coefficient!drag} for most bluff body\index{bluff body} shapes is a constant independent of Reynolds number, and it is generally obtained empirically. Table \ref{tab:Cd} shows the drag coefficients\index{coefficient!drag} for various two- and three-dimensional shapes. The shape clearly has a large impact, with $C_D$ varying from $1.17$ for a cylinder and $2.30$ for a forward facing cup (empty hemi-cylinder), which coincides well with the wake\index{wake} sizes shown in figure \ref{fig:dragWake}. Generally the drag decreases by around half for three-dimensional shapes of similar cross-section. As designers, we see that for quasi-steady flow (low reduced frequencies) we can maximize thrust by increasing the speed and size of the propulsor, and optimizing its shape. 

\begin{figure}
\centering
\includegraphics[width=0.85\textwidth]{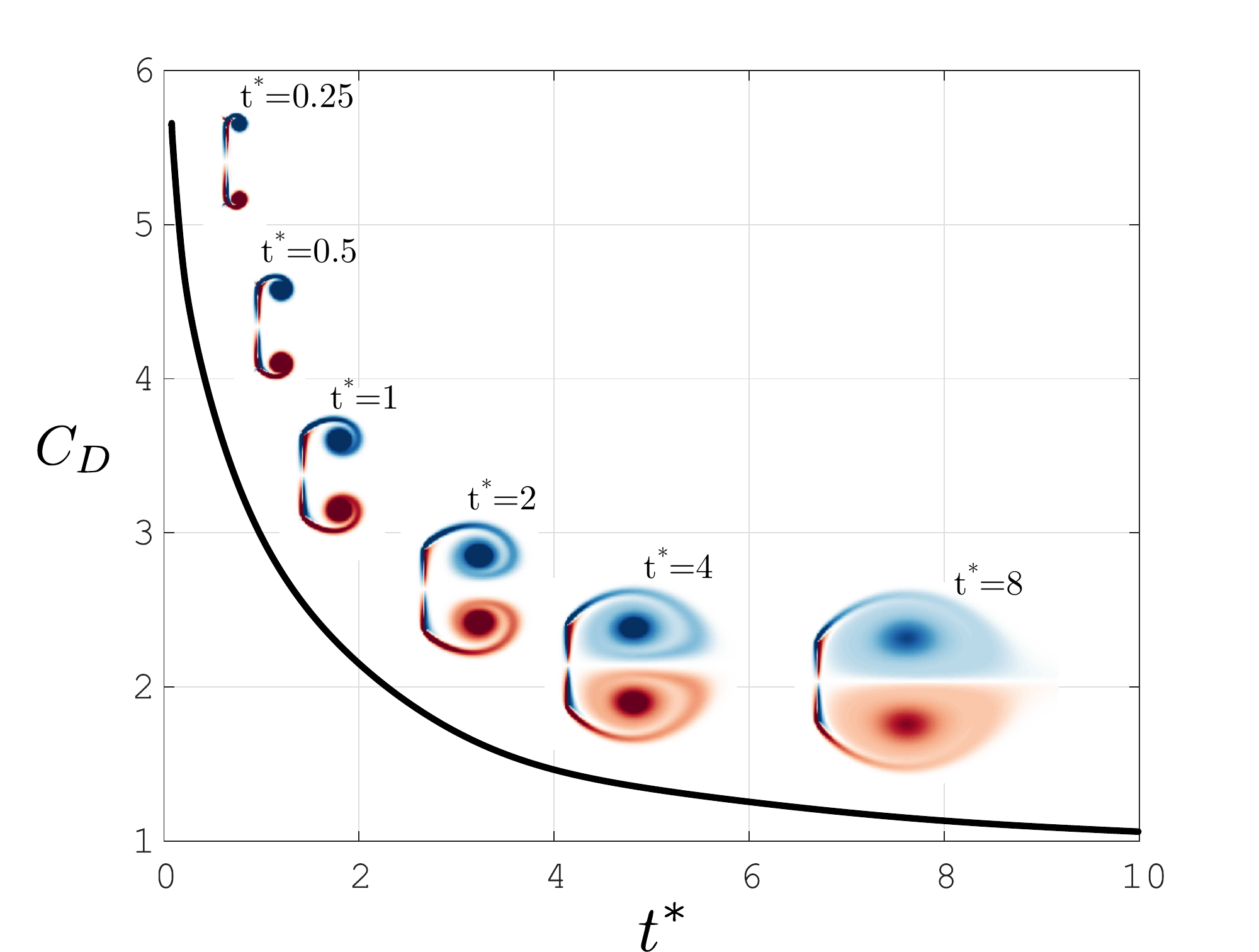}
\caption{Drag coefficient\index{coefficient!drag} for a impulsively started flat plate where the flow velocity is normal to the plate at Reynolds number $Re=500$. Inlaid graphics represent the qualitative vorticity field at six points in time. Simulations using immersed boundary code \cite{taira2007, colonius2008}.}
\label{fig:flatPlate}
\end{figure}

If we now consider unsteady flows, we can identify additional ways to increase thrust. Figure \ref{fig:flatPlate} shows a simulation of the drag coefficient\index{coefficient!drag} and wake of an impulsively started two-dimensional flat plate at Reynolds number $Re=500$. Here, we non-dimensionalize the time by the time it would take a fluid particle to travel a distance $X$, where $X$ is the height of the plate, thus $t^*=t\,(U_p/X)$. There is a dramatic decrease in drag between $0\leq t^*\leq 10$ as the vortices that shed from the plate edges grow to their maximum size. This means that if one could periodically and impulsively use a drag-based propulsor\index{thrust!drag-based} to generate thrust and limit the development time to $t^* \leq 2$, on average the thrust production would be four times higher than that produced by a quasi-steady drag based propulsor.  Although the Reynolds number\index{Reynolds number} of these simulations is relatively small, the broad conclusions will carry over to higher Reynolds numbers since, for a thin plate normal to a flow, there can be no net viscous drag\index{drag!viscous drag} in the flow direction, thus the form drag\index{drag!pressure/form drag} dominates.

\begin{figure}
\centering
\includegraphics[width=0.5\textwidth]{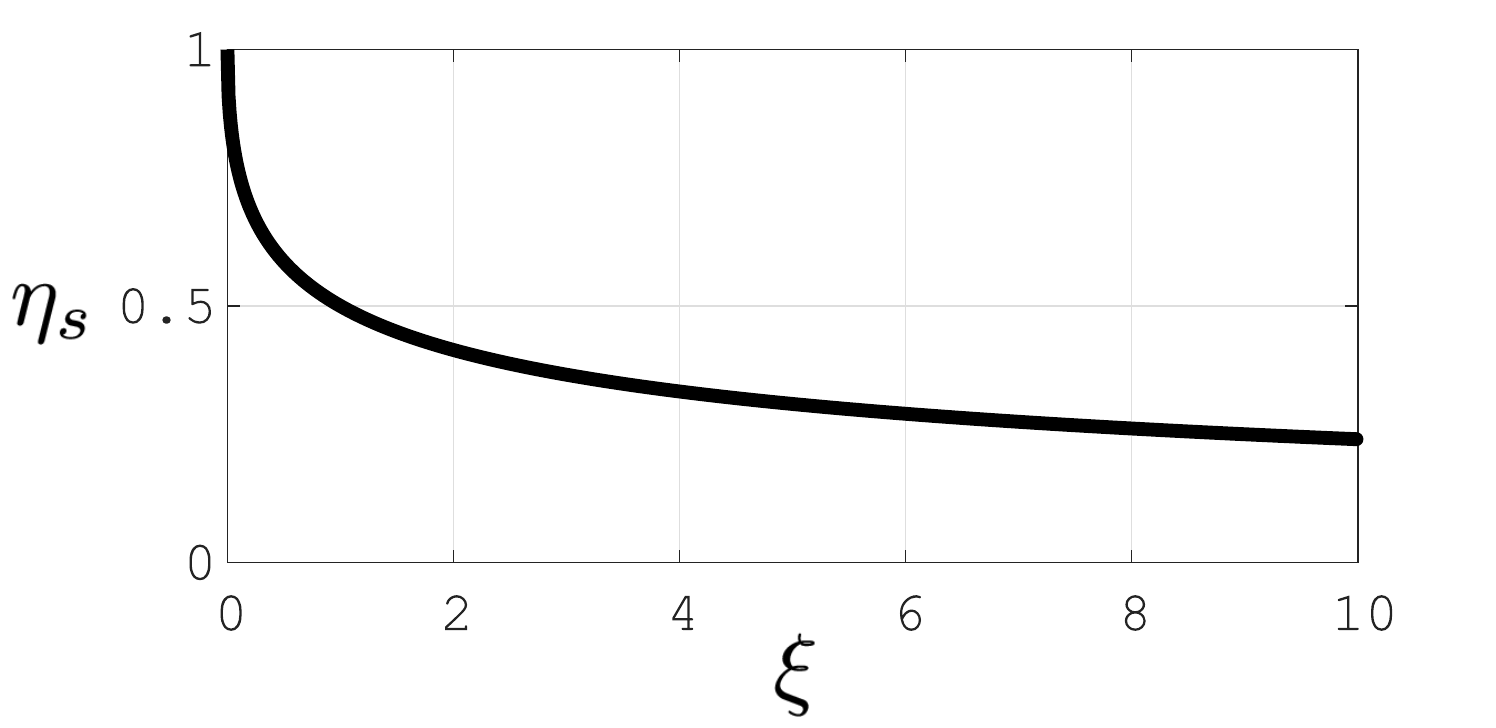}
\caption{Drag-based propulsion system efficiency.  Here,  $\xi=C_{D_b}A_b/(C_{D_p}A_p)$.}
\label{fig:dragEta}
\end{figure}

The efficiency\index{efficiency!swimming} of the drag-based propulsor cannot be considered in isolation as we did for lift-based\index{thrust!lift-based} propulsors, because efficiency is the ratio of an output to an input, and there is no clear output for a drag-based propulsor. However, we can define an efficiency if we include the entire vehicle. Consider a human paddling a canoe at constant velocity (for details see \cite{cabrera2006}). We define a system including the boat and the paddle but not the human. An energy balance\index{energy balance} of this system is
\begin{equation}
W_{ext} = \Delta E_K + \Delta E_P + W_{int},
\end{equation}
where $\Delta E_K$ and $\Delta E_P$ are the kinetic and potential energy changes of the system which are zero for constant velocity and unchanging height, and $W_{int}$ is the internal work of the system which we will neglect. The external work, $W_{ext}$ consists of the work done by the human on the paddle and the boat ($W_h$), the work done by the water on the paddle ($W_p$), and the work done by the water on the boat ($W_b$). The work of the human balances the resistance of the water, thus
\begin{equation}
W_h - W_p - W_b = 0
\end{equation}
and we can define a system efficiency that compares the useful output energy (boat moving forward) to the input energy (human)
\begin{equation}
\eta_s = \frac{W_b}{W_h} = \frac{W_b}{W_p+W_b}.
\end{equation}
If the work is constant, we can substitute power for work, and  power is given by the product of force and velocity, and so the efficiency is given by
\begin{equation}
\eta_s = \frac{F_b\,U_b}{F_p\,(U_p-U_b)+F_b\,U_b},
\end{equation}
where $F_b$ and $F_p$ are the drag on the boat and paddle, $U_b$ is the velocity of the boat, and $U_p$ is the velocity of the paddle relative to the boat. If the velocity is constant we have a balance of forces, $F_b=F_p$, and the efficiency\index{efficiency!swimming} just becomes a velocity ratio
\begin{equation}
\eta_s = \frac{U_b}{U_p}.
\label{eqn:dragEfficiency}
\end{equation}
From equation \ref{eqn:drag} combined with our balance of forces we obtain
\begin{equation}
U_b^2\, A_b\,C_{D_b}=(U_p-U_b)^2\, A_p\,C_{D_p}.
\end{equation}
Solving for $U_p$ in terms of $U_b$, applying the constraint that $U_p > U_b$ because to produce thrust the paddle must move faster than the boat, and using equation \ref{eqn:dragEfficiency} gives 
\begin{equation}
\eta_s = \frac{1}{1+\sqrt{\xi}},
\end{equation}
where $\xi$ is the ratio $C_{D_b}A_b/(C_{D_p}A_p)$. This result, obtained for paddling a canoe, can be used for any vehicle propelled using drag-based propulsor.  The theoretical system efficiency\index{efficiency!swimming} $\eta_s$ is shown in figure \ref{fig:dragEta} as a function of $\xi$, which indicates that for a high efficiency drag-based propulsion system it is best to have a large and high-drag propulsor and a small and low-drag\index{drag} vehicle. Interestingly, these are all things that also result in higher thrust and speed, thus coupling high thrust and efficiency. It is clear that drag-based propulsors are not inherently inefficient, and with proper design could be competitive with other propulsors.

\subsubsection{\emph{Concept design}}\index{underwater vehicle}\index{concept design}

Our concept vehicle, like most animals that utilize drag-based propulsion, will be amphibious. It can swim fully submerged, at the water surface, or traverse on land. It will be suitable for tasks that require versatility and large payloads.

Two tank-like tracks will be used to facilitate motion on both land an water. Inside the tracks are spring-loaded half-cylinders individually driven via solenoid, and can be impulsively exposed to the surrounding fluid. This maximizes our available thrust for drag-based\index{thrust!drag-based} propulsion, with the highest drag shape possible being periodically impulsively started. The spacing of the propulsors will promote drag via propulsor interaction, much like the interaction of schooling fish or flocking birds. The half-cylinders will only be exposed on a single side of each track which will allow water travel on one side and land travel on the other (requiring the vehicle to flip in transition). Note, if the vehicle was designed for submerged water travel only, the tracks could be turned on their side and shrouded to expose only the propulsors to be more hydrodynamic.

The central body will carry a dry payload and trap enough air for neutral buoyancy. Control surfaces will steer the vehicle during water travel.

\begin{figure}
\centering
\includegraphics[width=0.75\textwidth]{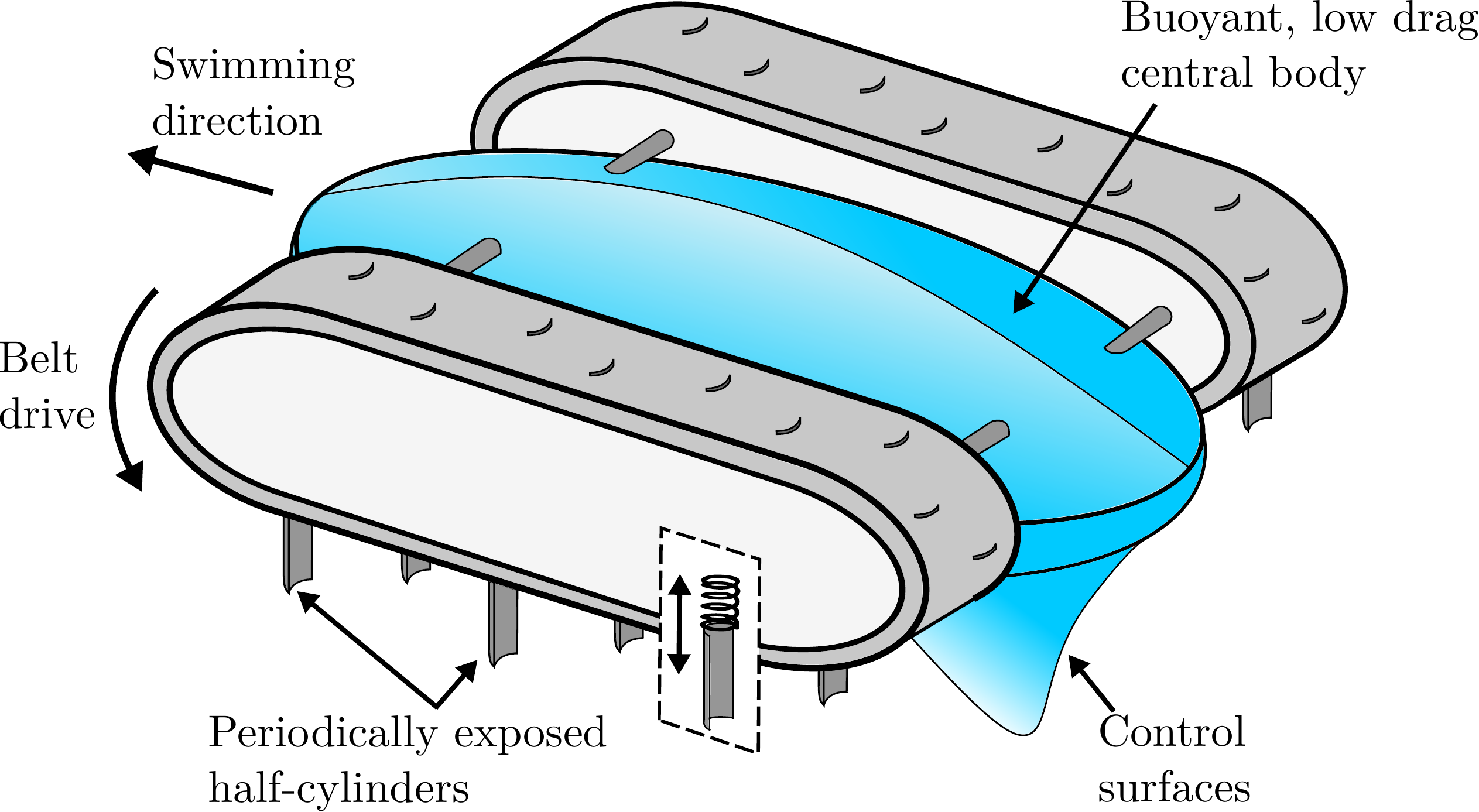}
\caption{Concept design\index{concept design} of an amphibious vehicle that uses drag-based propulsion\index{thrust!drag-based}, with open half-cylinders periodically exposed to the flow for limited time. The vehicle can be flipped to travel on the ground like a tank.}
\label{fig:concept3}
\end{figure}

\subsection{Concluding remarks}

We have attempted to provide an informative and intriguing summary of how aquatic swimmers can inspire biological propulsion systems. We identified three major groups of aquatic swimmers for inspiration. First were oscillatory and undulatory swimmers (for example, tuna\index{swimming animals!tuna} or manta ray\index{swimming animals!ray}), that periodically oscillate their flukes\index{fluke}, or median and caudal fins\index{caudal fin} to create lift-based\index{thrust!lift-based} and added mass\index{added mass} propulsion. We showed that for high thrust and efficiency\index{efficiency!swimming}, pitching\index{pitch} and heaving\index{heave} motions must be combined at a phase offset of about $270^\circ$, a limitation set by the need to reduce the peak angle of attack\index{angle of attack} of the motion. Adding flexibility\index{flexibility} was paramount to further increasing the thrust and efficiency of the propulsor.

Second were pulsatile swimmers (for example, jellyfish\index{swimming animals!jellyfish}) that create unsteady jets\index{jet} to inject momentum into the flow to produce thrust. The stroke ratio\index{stroke ratio}, $l/d_o$, was identified as one of the key governing parameters of performance.  A minimum stroke ratio of $\approx1$ is required to ensure the formation of a jet\index{jet}, otherwise the pulse of the previous stroke would be sucked back into the orifice. A maximum stroke ratio of $\approx4$ was required to prohibit the formation of a trailing jet\index{jet}, which negatively impacts the mean thrust of each pulse.

Lastly, drag-based swimmers  (for example, turtle\index{swimming animals!turtle}) that create thrust\index{thrust!drag-based} from the pressure-drag produced by moving a bluff body\index{bluff body} through the water. The propulsor shape is critical in maximizing thrust for a given size and velocity.  Quasi-two-dimensional shapes produced nearly twice as much drag as low aspect ratio three-dimensional shapes. Also, by limiting the flow development time from an impulsive start, the average thrust could be significantly increased. Drag-based propulsors appear to offer some previously unexplored potential for fast and efficient swimming. 

We hope that our proposed concept vehicles, based on what we currently know about biological propulsors, provide inspiration to future engineers and scientists to explore the design of future underwater vehicles\index{underwater vehicle}. 

This work was supported by ONR Grant N00014-14-1-0533 (Program Manager Robert Brizzolara). 

\printindex 

\bibliographystyle{plain}

\end{document}